\documentclass[twocolumn,amsmath,amssymb,a4paper,prb,superscriptaddress,floatfix]{revtex4-1}
\usepackage[dvipdfmx]{graphicx}
\usepackage{natbib}
\usepackage{multirow}
\usepackage{amsmath}
\usepackage{bm}
\usepackage{mathrsfs}
\usepackage{mathtools}
\usepackage{physics}
\usepackage{url}
\usepackage{xcolor}
\usepackage[normalem]{ulem}
\usepackage{hyperref}
\DeclareMathOperator*{\argmin}{arg\,min}

\begin{document}
\title{Implementation strategies in phonopy and phono3py}
\author{Atsushi Togo}
\email{togo.atsushi@gmail.com}
\affiliation{Research and Services Division of
  Materials Data and Integrated System, National Institute for Materials
  Science, Tsukuba, Ibaraki 305-0047, Japan}
\affiliation{Center for Elements Strategy Initiative for Structural
  Materials, Kyoto University, Sakyo, Kyoto 606-8501, Japan}
\author{Laurent Chaput}
\affiliation{Universit\'{e} de Lorraine, LEMTA, Centre National de la Recherche
Scientifique, Unit\'{e} Mixte de Recherche 7563, BP 70239, 54506 Vand\oe uvre
Cedex, france}
\author{Terumasa Tadano}
\affiliation{Research Center for Magnetic and Spintronic Materials,
  National Institute for Materials Science, Tsukuba, Ibaraki 305-0047,
  Japan}
\author{Isao Tanaka}
\affiliation{Center for Elements Strategy Initiative for Structural
  Materials, Kyoto University, Sakyo, Kyoto 606-8501, Japan}
\affiliation{Department of Materials Science and
  Engineering, Kyoto University, Sakyo, Kyoto 606-8501, Japan}
\affiliation{Nanostructures Research Laboratory, Japan Fine Ceramics
  Center, Atsuta, Nagoya 456-8587, Japan}

\begin{abstract}
  Scientific simulation codes are public property sustained by
  the community. Modern technology allows anyone to join scientific
  software projects, from anywhere, remotely via the internet.

  The phonopy and phono3py codes are  widely used open source phonon calculation
  codes. This review describes a collection of computational methods and
  techniques as implemented in these codes and shows their implementation
  strategies as a whole, aiming to be useful for the community. Some of the
  techniques presented here are not limited to phonon calculations and may
  therefore  be useful in other area of condensed matter physics.
\end{abstract}
\maketitle
\tableofcontents
\section{Introduction}
\label{sec:introduction}

Supported by an exponential growth of computer power, computer simulations and
scientific software development are essential for condensed matter physics and
materials science. Results obtained from computer simulations are getting more
realistic and scientific software development more robust.

Since many scientific codes are distributed under open-source
licenses, it is easy to start using scientific software and perform
computations. Participation in scientific software development has also become
easier with the availability of open-source compilers and source code editors.
Collaborative development by members at different worksites or locations, i.e.,
\textit{distributed} development, became quite popular for scientific software these
days. Anyone can contribute to an open-source software project, and the
contributions are usually united on internet hosting services for software
development. Documentation of the scientific software is an important part of
the software development, not only for users but also for scientific software
developers. For users, it explains how to use the software. For developers, it
describes how to read and write the code.

Development of key software can be a project that lasts for many years. Once a
key code is developed, it may be used for an unexpectedly long time. However,
computer architectures are constantly evolving. The way to achieve high
performance computing may therefore change drastically in a relatively short
time, on the order of a decade. For example, it is always required to follow the
increase in the number of cores in processors. Indeed, algorithms and data
structures must be optimized for a concurrent use of many cores in a multi-core
processor. Increasing memory space usually allows more flexible software design.
To adapt the code to those changes, the documentation is important to share the
meaning of the software design choices among users and developers.

The phonopy and phono3py codes are scientific software developed to perform
phonon calculations. A variety of physical properties can be calculated using
them: the phonon spectrum, the dynamical structure factor, and the lattice
thermal conductivity, to mention just a few. The computational method is based
on the supercell approach. In the computer implementations, various numerical
methods and techniques are employed. Some of those implemented in the phonopy
and phono3py codes are covered in this review. Our motivation to write this
review is to provide essential information to understand the codes in depth and
to invite scientific software developers in the phonopy and phono3py projects.

Although this review is written targeting scientific software developers, it may
be also useful for expert users of the phonopy and phono3py codes. The
computational methods and techniques underlying the phonon calculations are
described as implemented in the phonopy and phono3py codes.

In Sec.~\ref{sec:crystal-symmetry}, the representation of the crystal structure
and the crystal symmetry are presented in a crystallographic way. The reciprocal
space of the crystal and the crystal symmetry in the reciprocal space are also
described. Then, the phonon coordinates are introduced. The concept of Brillouin
zone (BZ) required for the phonon calculation is briefly defined.

In Sec.~\ref{seq:coordinates-supercell-approach}, the geometry of the supercell
structure model is described. The supercell geometry is associated to the
primitive cell by an integer matrix and how to construct the supercell geometry
using the integer matrix is explained.

In Sec.~\ref{sec:fc-to-dynmat-dynmat-to-fc}, the transformation between the
supercell force constants and the dynamical matrices is presented. Since the
phonopy and phono3py codes employ the supercell approach, the force constants
elements are limited to within the supercell. The transformation therefore needs
to be performed with special care. Such details are given in this section.

In Sec.~\ref{seq:non-analytical-term-correction}, the treatment of long range
dipole-dipole interaction in the dynamical matrix, as implemented in the phonopy
code, is explained. The implementation is mainly based on
Ref.~\onlinecite{Gonze-1997}, and its application to the supercell approach is
reviewed.

In Sec.~\ref{sec:regular-grids}, the traditional and generalized regular grids
used to sample reciprocal space discretely are defined. The symmetry treatment
of those grid points, used in the phonopy and phono3py codes, is also presented.

In Sec.~\ref{sec:tetrahedron-method}, a linear tetrahedron method is described
using the formulation as implemented in the phonopy and phono3py codes. The
implementation is mainly based on Refs.~\onlinecite{MacDonald-tetrahedron-1979}
and \onlinecite{Blochl-tetrahedron-1994}. The method is then applied to three
phonon scattering.

In Sec.~\ref{seq:random-displacement}, a scheme to generate random atomic
displacements in the supercell at finite temperatures is presented. This is
achieved by superpositions of displacements corresponding to randomly sampled
harmonic oscillators.

In Sec.~\ref{seq:phonon-band-unfolding}, a band unfolding technique implemented
in the phonopy code following Ref.~\onlinecite{Allen-band-unfolding} is
explained. The formulation specific for the phonon calculation with the
supercell approach is provided.

\section{Crystal and symmetry}
\label{sec:crystal-symmetry}

\subsection{Crystal structure}
\label{sec:crystal-structure}

\begin{figure}[ht]
  \begin{center}
    \includegraphics[width=0.8\linewidth]{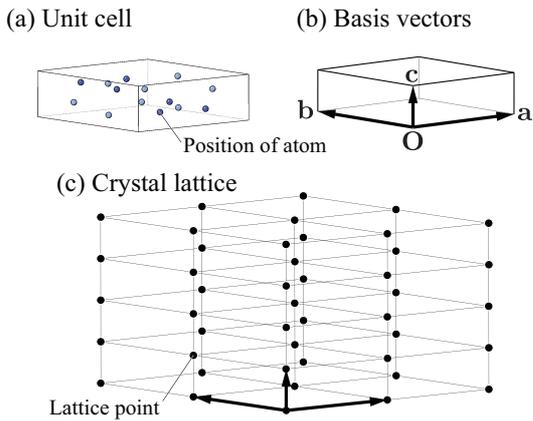}
    \caption{ Crystal structure model of $\beta$-Si$_3$N$_4$. See also
      Fig.~\ref{fig:spacegroup-operation} for the space group symmetry. (a)
      Unit cell containing atoms. (b) Basis vectors of the unit cell. (c) Crystal
      lattice. The filled circle symbols show the lattice points. In this model,
      the unit cell is assumed to be primitive. \label{fig:crystal-structure}}
  \end{center}
\end{figure}

As shown in Fig.~\ref{fig:crystal-structure}, a crystal model is defined by a
collection of unit cells periodically repeated in the three directions of space.
Each unit cell contains atoms. We normally choose a conventional set of basis
vectors to span the unit cell, $(\mathbf{a}, \mathbf{b}, \mathbf{c})$, so that
it represents naturally the crystal symmetry along with the choice of origin of
atomic positions.\cite{ITA} Positions of atoms are represented either in
Cartesian coordinates or in crystallographic coordinates. Denoting the
crystallographic coordinates of an atom as
$\boldsymbol{x}=(x_1,x_2,x_3)^\intercal $, its position $\mathbf{R}$ is given by
\begin{align}
  \mathbf{R} & = x_1 \mathbf{a} + x_2 \mathbf{b} + x_3 \mathbf{c}
  \nonumber                                                            \\
             & = (\mathbf{a}, \mathbf{b},  \mathbf{c})
  \begin{pmatrix} x_{1} \\  x_{2} \\  x_{3} \end{pmatrix}
  \nonumber                                                            \\
             & = (\mathbf{a}, \mathbf{b},  \mathbf{c}) \boldsymbol{x}.
\end{align}
In the above equation, if the basis vectors are represented using column vectors
containing their Cartesian components, then $ (\mathbf{a}, \mathbf{b},
  \mathbf{c})$ becomes a matrix, and $\mathbf{R}$ a column vector containing the
Cartesian coordinates of the atomic position vector,
\begin{align}
  \begin{pmatrix} \mathbf{R}_x \\  \mathbf{R}_y \\  \mathbf{R}_z \end{pmatrix} =
  \begin{pmatrix} \mathbf{a}_x & \mathbf{b}_x & \mathbf{c}_x \\
                \mathbf{a}_y & \mathbf{b}_y & \mathbf{c}_y \\
                \mathbf{a}_z & \mathbf{b}_z & \mathbf{c}_z\end{pmatrix}
  \begin{pmatrix} x_{1} \\  x_{2} \\  x_{3} \end{pmatrix}.
\end{align}
In the following we will use both representations. Depending on context, $
  (\mathbf{a}, \mathbf{b},  \mathbf{c})$ can therefore be considered as a row
vector containing three elements of a vector space, or a 3 by 3 matrix.

It is assumed that elements of $\boldsymbol{x}$ are in the interval $[0, 1)$ to
describe the structure. However it is often the case that we have to compare two
positions of atoms which may be in different unit cells, possibly after a space
group operation. For example, to identify two coordinates $\boldsymbol{x}$ and
$\boldsymbol{x}'$ which correspond to the same location within the unit cell,
but are possibly shifted by a lattice vector, it is convenient to bring each
element of the difference $\Delta\boldsymbol{x} = \boldsymbol{x}'
  -\boldsymbol{x}$ into the interval $[-0.5, 0.5)$ and then confirm that the
length of $\Delta\boldsymbol{x}$ is smaller than a tolerance $\epsilon$. This is
implemented in the phonopy and phono3py codes everywhere by rounding components
of $\Delta\boldsymbol{x}$ to the nearest integer
($\text{nint}(\Delta\boldsymbol{x})$) and checking
$\left|(\mathbf{a},
  \mathbf{b}, \mathbf{c}) [\Delta\boldsymbol{x} -
      \text{nint}(\Delta\boldsymbol{x})] \right| < \epsilon$.

\subsection{Space group operation}

\begin{figure}[ht]
  \begin{center}
    \includegraphics[width=1.0\linewidth]{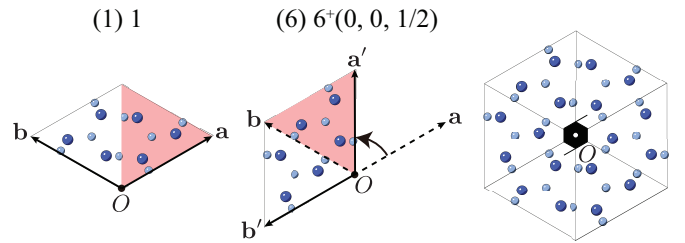}
    \caption{Symmetry operations applied on the unit cell of $\beta$-Si$_3$N$_4$
      whose space-group type is $P6_3/m$ (No. 176). The left and middle figures
      illustrate identity and 6-fold screw axis operations. Their notations
      follow the international tables for crystallography volume A.\cite{ITA}
      The right figure shows the crystal structure viewed from the top
      ($z$-axis) that depicts the 6-fold screw axis passing through the origin
      $O$. \label{fig:spacegroup-operation} }
  \end{center}
\end{figure}

Various symmetries of crystals and phonons are used in the phonopy and phono3py
codes. The most important one is the space group symmetry. A space group
operation is composed of a  (proper or improper) rotation $\mathcal{S}$ and a
translation $\tau$. It is often written by a composite notation
$(\mathcal{S},\tau)$ which sends the position $\mathbf{R}$ to
\begin{align}
  \mathbf{R}'=(\mathcal{S},\tau)\mathbf{R}\equiv\mathcal{S}\mathbf{R}+\tau.
\end{align}
Using  crystallographic coordinates this is written as
\begin{align}
  (\mathbf{a}, \mathbf{b}, \mathbf{c}) \boldsymbol{x}' =
  \mathcal{S} (\mathbf{a}, \mathbf{b}, \mathbf{c})
  \boldsymbol{x}+ (\mathbf{a}, \mathbf{b},  \mathbf{c}) \boldsymbol{w}
\end{align}
or
\begin{align}
  \boldsymbol{x}' & =(\mathbf{a}, \mathbf{b}, \mathbf{c})^{-1} \mathcal{S}
  (\mathbf{a}, \mathbf{b},  \mathbf{c}) \boldsymbol{x}+  \boldsymbol{w}
  \nonumber                                                                \\
                  & \equiv \boldsymbol{S} \boldsymbol{x} + \boldsymbol{w},
\end{align}
where $\boldsymbol{S}$ can be shown to be a unimodular
matrix. $\boldsymbol{w}$
contains the components of the translation vector in the crystallographic
basis. If Cartesian coordinates were used instead, the matrix representing the
rotation would be orthogonal. Representing the space group by
$\mathbb{S}=\{(\mathcal{S},\tau)\}$, the crystallographic point group is given
by $\mathbb{P}=\{\mathcal{S}|(\mathcal{S},\tau) \in \mathbb{S}\}$. An example of
space group operation is shown in Fig.~\ref{fig:spacegroup-operation}. The
crystal structure overlaps with itself after a rotation of 60$^\circ$ along $z$
axis and a $+\frac{1}{2}\mathbf{c}$ shift. Therefore
\begin{align}
  \boldsymbol{S}= \begin{pmatrix}
                    1 & \bar{1} & 0 \\
                    1 & 0       & 0 \\
                    0 & 0       & 1
                  \end{pmatrix},
  \boldsymbol{w}=
  \begin{pmatrix} 0 \\ 0 \\ 1/2 \end{pmatrix},
\end{align}
must be a space group operation. It means that any atom in the unit cell at
point $\boldsymbol{x}$ is sent by the space group operation to a new point
\begin{align}
  \boldsymbol{x}' = \begin{pmatrix}
                      1 & \bar{1} & 0 \\
                      1 & 0       & 0 \\
                      0 & 0       & 1
                    \end{pmatrix} \boldsymbol{x} + \begin{pmatrix} 0 \\ 0 \\ 1/2 \end{pmatrix}.
\end{align}
The new point $\boldsymbol{x}'$ can be located out of the original unit cell,
but an atom with the same atomic type must be found at this new location.

\subsection{Primitive cell}

\begin{figure}[ht]
  \begin{center}
    \includegraphics[width=0.7\linewidth]{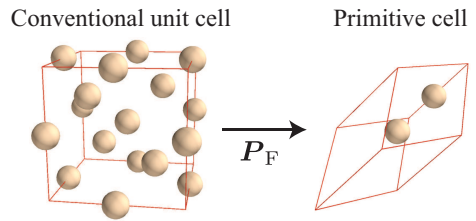}
    \caption{ Conventional unit cell and primitive cell of silicon. The space
      group type is $Fd\bar{3}m$ (No.~227). Since the conventional unit cell
      contains four lattice points with eight atoms,
      the primitive cell is chosen to have two
      atoms by $\det(\boldsymbol{P}_F)=1/4$.\label{fig:primitive-cell} }
  \end{center}
\end{figure}

In most cases, a unit cell as described in Sec.~\ref{sec:crystal-structure} is
used as the input crystal structure model of a phonon calculation. The most
plausible choice of unit cell is either a primitive cell or a conventional unit
cell. The former represents the minimum unit of periodicity of a crystal
lattice. It is therefore the one used for Fourier expansions and is necessary
for a reciprocal space representation of phonon properties. The Bloch theorem
evidenced this lattice translational symmetry. The latter follows the
crystallographic convention. Although it provides intuitive shapes of the unit
cells (cubic, hexagonal, etc), it may contain several lattice points.
Fig.~\ref{fig:primitive-cell} shows the conventional unit cell and a primitive
cell for silicon crystallized within the face-centred-cubic structure.

The conventional unit cell structure is uniquely defined to follow the
crystallographic convention,\cite{ITA} within the freedom
due to Euclidean normalizer.\cite{ITA} On the other hand, the choice of
primitive cell is not unique. This is inconvenient for a systematic handling of
crystal structure models in phonon calculations. Therefore, phonopy suggests
predefined transformation matrix $\boldsymbol{P}_\text{prim}$ from the
conventional unit cell to the primitive cell. $\boldsymbol{P}_\text{prim}$ is
used as
\begin{align}
  \label{eq:conv-to-prim-matrix}
  (\mathbf{a}_\text{p},
  \mathbf{b}_\text{p}, \mathbf{c}_\text{p}) = (\mathbf{a}_\text{c},
  \mathbf{b}_\text{c}, \mathbf{c}_\text{c})\boldsymbol{P}_\text{prim},
\end{align}
where $(\mathbf{a}_\text{p}, \mathbf{b}_\text{p}, \mathbf{c}_\text{p})$ are the
basis vectors of the primitive cell and $(\mathbf{a}_\text{c},
  \mathbf{b}_\text{c}, \mathbf{c}_\text{c})$ those of the conventional unit cell.
The choices of the transformation matrices used in the phonopy
and phono3py codes are shown in Table \Ref{tab:trans-c2p}.
For example, the
transformation matrix used for silicon in Fig.~\ref{fig:primitive-cell} is
$\boldsymbol{P}_\text{prim}=\boldsymbol{P}_F$. These matrices are equivalent to
those presented in Table 2 of Ref.~\onlinecite{Aroyo-2014} although those are
given for the reciprocal space basis vectors.

\begin{table}
  \caption{\label{tab:trans-c2p} Choices of transformation matrices
    $\boldsymbol{P}_\text{prim}$ of Eq.~(\ref{eq:conv-to-prim-matrix}) used in the
    phonopy and phono3py codes. The subscripts $X$ of the matrices
    $\boldsymbol{P}_X$ indicate the centring types: $A$, $B$, $C$ for the base
    centring types, $I$ and $F$ for the body and face centring types,
    respectively, and $R$ for the (obverse) rhombohedral centring type. }
  \begin{center}
    \begin{tabular}{lll}
      $\boldsymbol{P}_A =
        \renewcommand*{\arraystretch}{1.3}
        \begin{pmatrix}
          1 & 0           & 0                 \\
          0 & \frac{1}{2} & \bar{\frac{1}{2}} \\
          0 & \frac{1}{2} & {\frac{1}{2}}     \\
        \end{pmatrix}
      $, &
      $\boldsymbol{P}_B =
        \renewcommand*{\arraystretch}{1.3}
        \begin{pmatrix}
          \frac{1}{2} & 0 & \bar{\frac{1}{2}} \\
          0           & 1 & 0                 \\
          \frac{1}{2} & 0 & {\frac{1}{2}}     \\
        \end{pmatrix}
      $, &
      $\boldsymbol{P}_C =
        \renewcommand*{\arraystretch}{1.3}
        \begin{pmatrix}
          \frac{1}{2}       & \frac{1}{2} & 0 \\
          \bar{\frac{1}{2}} & \frac{1}{2} & 0 \\
          0                 & 0           & 1 \\
        \end{pmatrix}
      $,   \\ \\
      $\boldsymbol{P}_I =
        \renewcommand*{\arraystretch}{1.3}
        \begin{pmatrix}
          \bar{\frac{1}{2}} & \frac{1}{2}       & \frac{1}{2}       \\
          {\frac{1}{2}}     & \bar{\frac{1}{2}} & \frac{1}{2}       \\
          {\frac{1}{2}}     & \frac{1}{2}       & \bar{\frac{1}{2}} \\
        \end{pmatrix}
      $, &
      $\boldsymbol{P}_F =
        \renewcommand*{\arraystretch}{1.3}
        \begin{pmatrix}
          0             & \frac{1}{2} & \frac{1}{2} \\
          {\frac{1}{2}} & 0           & \frac{1}{2} \\
          {\frac{1}{2}} & \frac{1}{2} & 0           \\
        \end{pmatrix}
      $, &
      $\boldsymbol{P}_R =
        \renewcommand*{\arraystretch}{1.3}
        \begin{pmatrix}
          \frac{2}{3} & \bar{\frac{1}{3}} & \bar{\frac{1}{3}} \\
          \frac{1}{3} & \frac{1}{3}       & \bar{\frac{2}{3}} \\
          \frac{1}{3} & \frac{1}{3}       & \frac{1}{3}       \\
        \end{pmatrix}
      $.   \\ \\
    \end{tabular}
  \end{center}
\end{table}

The crystal lattice is defined as the set of integer linear combinations of
primitive lattice vectors. A lattice vector $\mathbf{R}_l$  can therefore be
written as
\begin{align}
  \mathbf{R}_l =
  l_1\mathbf{a}_\text{p}
  + l_2\mathbf{b}_\text{p} + l_3\mathbf{c}_\text{p}=
  (\mathbf{a}_\text{p}, \mathbf{b}_\text{p}, \mathbf{c}_\text{p}) \boldsymbol{l}  \;\; (l_i \in \mathbb{Z}).
\end{align}
In Fig.~\ref{fig:crystal-structure}, if the unit cell is assumed to be
primitive, then Fig.~\ref{fig:crystal-structure} (c) shows the crystal lattice.
The lattice point may then be chosen to coincide with the origin of each unit
cell. Sometimes it is also useful to regard the crystal lattice as generated
using the lattice translations of the space group,
$(\boldsymbol{I},\boldsymbol{t})$, where $\boldsymbol{I}$ and $\boldsymbol{t}$
denote the identity operation and lattice translation, respectively.

\subsection{Reciprocal space}

The crystal structure, and the symmetry explained above, are defined in direct
(real) space. In this section, crystallography is briefly summarized in
reciprocal space. This is quite useful for our purpose since phonons are
described in reciprocal space using wave vectors. It is convenient to introduce a
reciprocal lattice whose basis vectors $(\mathbf{a}_\text{p}^*,
  \mathbf{b}_\text{p}^*, \mathbf{c}_\text{p}^*)$ are solutions of the equation
\begin{align}
  (\mathbf{a}_\text{p}, \mathbf{b}_\text{p},  \mathbf{c}_\text{p})^{\intercal}  (\mathbf{a}_\text{p}^*, \mathbf{b}_\text{p}^*, \mathbf{c}_\text{p}^*)      = 2 \pi \boldsymbol{1}.
  \label{eq:rcell}
\end{align}
This equation can be solved to give
\begin{align}
  \mathbf{a}_\text{p}^* = 2\pi \frac{\mathbf{b}_\text{p} \times \mathbf{c}_\text{p}}{V_\text{c}}, \;\;
  \mathbf{b}_\text{p}^* = 2\pi \frac{\mathbf{c}_\text{p} \times \mathbf{a}_\text{p}}{V_\text{c}}, \;\;
  \mathbf{c}_\text{p}^* = 2\pi \frac{\mathbf{a}_\text{p} \times \mathbf{b}_\text{p}}{V_\text{c}},
\end{align}
where $V_\text{c} = \mathbf{a}_\text{p} \cdot (\mathbf{b}_\text{p} \times
  \mathbf{c}_\text{p}) = \mathbf{b}_\text{p} \cdot (\mathbf{c}_\text{p} \times
  \mathbf{a}_\text{p}) = \mathbf{c}_\text{p} \cdot (\mathbf{a}_\text{p} \times
  \mathbf{b}_\text{p})$. The reciprocal lattice points are then obtained from the
integer linear combinations of reciprocal basis vectors,
\begin{align}
  \mathbf{G} = G_1\mathbf{a}_\text{p}^* + G_2\mathbf{b}_\text{p}^* + G_3\mathbf{c}_\text{p}^*=
  (\mathbf{a}_\text{p}^*, \mathbf{b}_\text{p}^*, \mathbf{c}_\text{p}^*) \boldsymbol{G}
  \;\; (G_i \in \mathbb{Z}).
\end{align}
Similarly to the atomic positions $\mathbf{R}$, wave vectors $\mathbf{q}$ are
defined from real linear combinations of the reciprocal lattice vectors,
\begin{align}
  \mathbf{q} =
  q_1\mathbf{a}_\text{p}^* + q_2\mathbf{b}_\text{p}^* +
  q_3\mathbf{c}_\text{p}^* =(\mathbf{a}_\text{p}^*,
  \mathbf{b}_\text{p}^*, \mathbf{c}_\text{p}^*) \boldsymbol{q} .
\end{align}
If the $q_i$ are restricted to the interval $[0,1)$, a primitive cell of
reciprocal space is obtained.

If we consider a component of a Fourier expansion,
$f_{\mathbf{q}}(\mathbf{R})=e^{i \mathbf{q}\cdot \mathbf{R}}$, it is transformed
by an operation of the  crystallographic point group to the function
\begin{align}
  \mathcal{S} f_{\mathbf{q}}(\mathbf{R}) & =
  f_{\mathbf{q}}(\mathcal{S}^{-1}\mathbf{R})=
  e^{i \mathbf{q}\cdot \mathcal{S}^{-1} \mathbf{R}}=
  e^{i \mathcal{S}\mathbf{q}\cdot \mathcal{S}\mathcal{S}^{-1} \mathbf{R}}=
  e^{i \mathcal{S}\mathbf{q}\cdot  \mathbf{R}} \nonumber                            \\
                                         & = f_{\mathcal{S}\mathbf{q}}(\mathbf{R}).
\end{align}
For this reason, the image of a wave vector through an operation of the
crystallographic point group is defined to be
\begin{align}
  \mathbf{q}'=\mathcal{S}\mathbf{q} \Longleftrightarrow
  (\mathbf{a}_\text{p}^*, \mathbf{b}_\text{p}^*, \mathbf{c}_\text{p}^*)
  \boldsymbol{q}' & =
  \mathcal{S}
  (\mathbf{a}_\text{p}^*, \mathbf{b}_\text{p}^*, \mathbf{c}_\text{p}^*)
  \boldsymbol{q}
\end{align}
or
\begin{align}
  \boldsymbol{q}' & =
  (\mathbf{a}_\text{p}^*, \mathbf{b}_\text{p}^*, \mathbf{c}_\text{p}^*)^{-1}
  \mathcal{S}
  (\mathbf{a}_\text{p}^*, \mathbf{b}_\text{p}^*, \mathbf{c}_\text{p}^*)
  \boldsymbol{q}                                                                                  \\
                  & = (\mathbf{a}_\text{p}, \mathbf{b}_\text{p}, \mathbf{c}_\text{p})^{\intercal}
  \mathcal{S}
  (\mathbf{a}_\text{p}, \mathbf{b}_\text{p}, \mathbf{c}_\text{p})^{-\intercal}
  \boldsymbol{q}                                                                                  \\
                  & = \boldsymbol{S}^{-\intercal}\boldsymbol{q}. \label{qprime}
\end{align}

The set $\{\boldsymbol{S}^{-\intercal} \boldsymbol{q}\}$ with $\boldsymbol{S}$
in the crystallographic point group $\{\boldsymbol{S}\}$ is called the star of
$\mathbf{q}$.

\subsection{Phonon coordinates}
Atoms vibrate in the vicinity of their equilibrium positions in crystals.
Instantaneous and equilibrium positions of atoms in
crystals are denoted by $\mathbf{R}_{l\kappa}$ and $\mathbf{R}^0_{l\kappa}$,
respectively, where $l$ and $\kappa$ are used to label the lattice points and
atoms in the primitive cell of $l$, respectively. Displacements of atoms are
written as $\mathbf{u}_{l\kappa} = \mathbf{R}_{l\kappa} -
  \mathbf{R}^0_{l\kappa}$.

In this review, we define the dynamical matrix as~\cite{phonopy-phono3py}
\begin{align}
  \label{eq:dynamical-matrix}
  D_{{\kappa}\alpha,{\kappa}'\alpha'}(\mathbf{q}) =
  \frac{1}{\sqrt{m_{\kappa}m_{{\kappa}'}}} \sum_{l'}
  \Phi_{0{\kappa}\alpha,l'{\kappa}'\alpha'}
  e^{i{\mathbf{q} \cdot (\mathbf{R}^0_{l'{\kappa}'} - \mathbf{R}^0_{0{\kappa}})}}.
\end{align}
$\alpha$ and $m_\kappa$ denote the index of the Cartesian coordinates and the
mass of the atom $\kappa$, respectively.
$\Phi_{l{\kappa}\alpha,l'{\kappa}'\alpha'}$ are the harmonic force constants,
defined as the second derivative of the energy with respect to atomic
positions, and evaluated at the equilibrium positions.\cite{phonopy-phono3py}
Phonon frequency $\omega_{\mathbf{q}\nu}$ and eigenvector
$W_{\kappa\alpha}(\mathbf{q}\nu)$ are obtained as the solution of the eigenvalue
equation of the dynamical matrix in Eq.~(\ref{eq:dynamical-matrix}), which is
written as
\begin{align}
  \label{eq:phonon-eigenvalue-problem}
  \sum_{\kappa'\alpha'} D_{{\kappa}\alpha,{\kappa}'\alpha'}(\mathbf{q})
  W_{\kappa'\alpha'}(\mathbf{q}\nu)
  = \omega^2_{\mathbf{q}\nu} W_{\kappa\alpha}(\mathbf{q}\nu).
\end{align}
$\nu$ labels the phonon band index, and the composite index $\mathbf{q}\nu$ is
used to consider a phonon mode.

Eq.~(\ref{eq:phonon-eigenvalue-problem}) can also be written in matrix form as
\begin{align}
  \label{eq:phonon-eigenvalue-problem-matrix}
  \mathrm{D}(\mathbf{q}) \mathrm{W}(\mathbf{q}) = \mathrm{W}(\mathbf{q})
  \Omega^2(\mathbf{q}),
\end{align}
or, because $ \mathrm{D}(\mathbf{q})$ is hermitian,
\begin{align}
  \label{eq:dynamical-matrix-in-matrix}
  \mathrm{D}(\mathbf{q}) = \mathrm{W}(\mathbf{q})
  \Omega^2(\mathbf{q}) \mathrm{W}^\dagger(\mathbf{q}),
\end{align}
where $\Omega^2(\mathbf{q})$ is the diagonal matrix whose diagonal elements are
$\omega^2_{\mathbf{q}\nu}$.  Each column of $\mathrm{W}(\mathbf{q})$ contains an eigenvector $W_{\kappa\alpha}(\mathbf{q}\nu)$ corresponding to a different band index $\nu$. Elements of each column are ordered as
$(W_{\kappa_1 x}, W_{\kappa_1 y}, W_{\kappa_1 z}, \ldots, W_{\kappa_{n_\text{a}}
      x}, W_{\kappa_{n_\text{a}} y}, W_{\kappa_{n_\text{a}} z})$, where $n_\text{a}$
is the number of atoms in the primitive cell.

Because
$D_{{\kappa}\alpha,{\kappa}'\alpha'}(\mathbf{q}) =
  D^*_{{\kappa}\alpha,{\kappa}'\alpha'}(-\mathbf{q})$,
we have
\begin{align}
  \omega_{\mathbf{q\nu}} = \omega_{\mathbf{-q\nu}},
\end{align}
and we can choose
\begin{align}
  W_{\kappa\alpha}(\mathbf{q}\nu) = W^*_{\kappa\alpha}(-\mathbf{q}\nu).
\end{align}

Moreover eigenvalues and eigenvectors inherit symmetry properties of the force
constants. It can be shown~\cite{Maradudin-RMP-1968} that under a space group
operation $(\mathcal{S},\tau)$ the eigenvalues and eigenvectors transform as
\begin{align}
  \label{eq:eigenvector-rotation}
   & \omega_{\mathcal{S} \mathbf{q\nu}} = \omega_{\mathbf{q\nu}}, \\
   & \mathbf{W}_{\kappa'}(\mathcal{S} \mathbf{q}\nu)
  = \mathcal{S}
  \mathbf{W}_{\kappa}(\mathbf{q}\nu)
  e^{-i\mathcal{S} \mathbf{q}\cdot \tau},
\end{align}
where $\mathbf{W}_{\kappa}(\mathbf{q}\nu)$ represents the phonon eigenvector of
atom $\kappa$ in Cartesian coordinates, and $\kappa'$ is the image of atom
$\kappa $ under the space group operation.

An actual displacement can be described from a linear combination of phonon
eigenvectors. This defines the phonon coordinates $Q(\mathbf{q}\nu)$ as
\begin{align}
  \label{eq:displacement-operator}
  u_{l{\kappa}\alpha} & = \sum_{\mathbf{q}\nu}
  Q(\mathbf{q}\nu)
  \frac{1}{\sqrt{Nm_{\kappa}}} W_{\kappa\alpha}(\mathbf{q}\nu) e^{i\mathbf{q}
  \cdot \mathbf{R}^0_{l{\kappa}}}
  \\
                      & = \sum_{\mathbf{q}\nu}
  Q(\mathbf{q}\nu) u_{l{\kappa}\alpha}(\mathbf{q}\nu),
\end{align}
where $N$ is the number of lattice points in crystal.
$u_{l{\kappa}\alpha}(\mathbf{q}\nu)$ is defined for the purpose of convenience
in the following sections.
The previous equation can be inverted to give
\begin{align}
  \label{eq:displacement-operator-inv}
  Q(\mathbf{q}\nu)= \sum_{l{\kappa}\alpha}
  u_{l{\kappa}\alpha} \sqrt{\frac{m_{\kappa}}{N}}
  W^{*}_{\kappa\alpha}(\mathbf{q}\nu) e^{-i\mathbf{q}
  \cdot \mathbf{R}^0_{l{\kappa}}}.
\end{align}

\subsection{Brillouin zone}
Symmetry property of phonons in reciprocal space is best represented in the
BZ.\cite{Aroyo-2014, ITB, Maradudin-RMP-1968,
  Mathematical_Theory_of_Symmetry_in_Solids, Kronecker_Product_Tables_Vol1,
  DynamicalPropertiesOfSolidsVol1} In the phonopy and phono3py codes, the BZ is
defined as a Wigner-Seitz cell of the reciprocal lattice. To check if a
$\mathbf{q}$ point belongs to the BZ we proceed as follow. First the basis
vectors of the Niggli cell~\cite{Niggli-1928, Gruber-1973, Krivy-1976,
  Grosse-Kunstleve-Niggli-2004} are determined. The Niggli cell is a cell with the
shortest possible reciprocal basis vectors. Then using integer linear
combinations of those Niggli basis vectors, we search for the shortest
$|\mathbf{q}+ \mathbf{G}|$.  Finally $\mathbf{q} + \mathbf{G}$ is used as the
$\mathbf{q}$ point in the BZ.

Three $\mathbf{q}$ points are needed for the three phonon scattering considered
in the phono3py code. In case one or more of those three points is on the BZ
surface, we choose the translationally equivalent points on the BZ surface which
minimize $|\mathbf{q} + \mathbf{q}' + \mathbf{q}''|$.

\begin{figure}[ht]
  \begin{center}
    \includegraphics[width=0.65\linewidth]{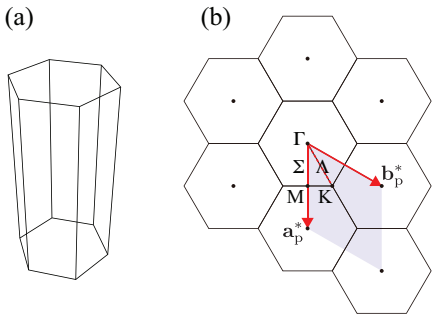}
    \caption{ Brillouin zone (BZ) of $\beta$-Si$_3$N$_4$. See also
      Figs.~\ref{fig:crystal-structure} and \ref{fig:spacegroup-operation} for the
      cell shape in direct space. (a) The first BZ. (b) BZs viewed from
      $\mathbf{c}_\text{p}^*$ axis direction. Symbols of the special points and
      paths follow the Bilbao crystallographic server.\cite{Aroyo-2014}
      \label{fig:hexagonal-BZ} }
  \end{center}
\end{figure}

As an example, the BZ of $\beta$-Si$_3$N$_4$ is presented in
Fig.~\ref{fig:hexagonal-BZ}. The basal plane has the hexagonal
shape and $\mathbf{c}_\text{p}^*$ is longer than $\mathbf{a}_\text{p}^*$ and
$\mathbf{b}_\text{p}^*$ because $\mathbf{c}_\text{p}$ is shorter than
$\mathbf{a}_\text{p}$ and $\mathbf{b}_\text{p}$. By definition,
$\mathbf{a}_\text{p}^*/2$, $\mathbf{b}_\text{p}^*/2$, and
$\mathbf{c}_\text{p}^*/2$ are located on the BZ surface. The high symmetry
points and paths in the BZ have special symbols as shown in
Fig.~\ref{fig:hexagonal-BZ} (b).\cite{Aroyo-2014}

\section{Geometry of supercell model}
\label{seq:coordinates-supercell-approach}

\subsection{Supercell construction}
\label{seq:supercell-construction}

In the phonopy and phono3py codes, the supercell approach is employed. The
supercell is defined by multiple primitive cells, so that the basis vectors of
the supercell $(\mathbf{a}_\text{s}, \mathbf{b}_\text{s}, \mathbf{c}_\text{s})$
can be represented as the image of the primitive cell basis vectors  through an
integer matrix $\boldsymbol{M}_{\text{p}\rightarrow\text{s}}$,
\begin{align}
  \label{eq:primitive-to-supercell}
  (\mathbf{a}_\text{s},
  \mathbf{b}_\text{s}, \mathbf{c}_\text{s}) =
  (\mathbf{a}_\text{p},
  \mathbf{b}_\text{p}, \mathbf{c}_\text{p})
  \boldsymbol{M}_{\text{p}\rightarrow\text{s}}.
\end{align}
Like the unit cell, the supercells are arranged on the supercell lattice.
The supercell lattice points are labeled by $L$, and located from the vectors $\mathbf{R}_L$.

\begin{figure}[ht]
  \begin{center}
    \includegraphics[width=0.9\linewidth]{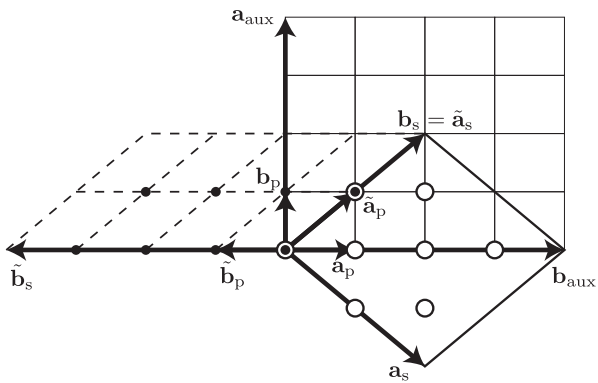}
    \caption{Lattices and basis vectors of primitive cell $(\mathbf{a}_\text{p},
        \mathbf{b}_\text{p})$, supercell $(\mathbf{a}_\text{s},
        \mathbf{b}_\text{s})$, new primitive cell
      $(\tilde{\mathbf{a}}_\text{p}, \tilde{\mathbf{b}}_\text{p})$, new
      supercell $(\tilde{\mathbf{a}}_\text{s},
        \tilde{\mathbf{b}}_\text{s})$, and auxiliary supercell
      $(\mathbf{a}_\text{aux}, \mathbf{b}_\text{aux})$. Circle and small
      filled circle symbols depict lattice points in the supercell and in the
      new supercell, respectively. The latter points can be brought by the
      supercell lattice translations to the former points uniquely.
      \label{fig:supercell-construction}
    }
  \end{center}
\end{figure}

The supercell model is constructed using
$\boldsymbol{M}_{\text{p}\rightarrow\text{s}}$. The lattice points
within the supercell are used to perform the
summation found in Eq.~(\ref{eq:dynamical-matrix}). These
lattice points can be elegantly obtained using the approach reported by  Hart
and Forcade.\cite{Gus-SNF-2008} The integer matrix $
  \boldsymbol{M}_{\text{p}\rightarrow\text{s}}$ is first reduced to a diagonal
integer matrix $\boldsymbol{D} = \boldsymbol{P}
  \boldsymbol{M_{\text{p}\rightarrow\text{s}}} \boldsymbol{Q}$ by the unimodular
matrices $\boldsymbol{P}$ and $\boldsymbol{Q}$, where we choose
$\det(\boldsymbol{P})=1$ and $\det(\boldsymbol{Q})=1$. This matrix decomposition
can be the Smith normal form (SNF), however, it is unnecessary to strictly
follow the definition of the SNF.

Using this transformation, Eq. \ref{eq:primitive-to-supercell} can be written
\begin{align}
  (\tilde{\mathbf{a}}_\text{s},
  \tilde{\mathbf{b}}_\text{s}, \tilde{\mathbf{c}}_\text{s}) =
  (\tilde{\mathbf{a}}_\text{p},
  \tilde{\mathbf{b}}_\text{p},\tilde{\mathbf{c}}_\text{p})
  \boldsymbol{D},
\end{align}
with
\begin{align}
   & (\tilde{\mathbf{a}}_\text{s},  \tilde{\mathbf{b}}_\text{s}, \tilde{\mathbf{c}}_\text{s}) =
  (\mathbf{a}_\text{s},  \mathbf{b}_\text{s}, \mathbf{c}_\text{s})   \boldsymbol{Q},            \\
   & (\tilde{\mathbf{a}}_\text{p}, \tilde{\mathbf{b}}_\text{p},\tilde{\mathbf{c}}_\text{p}) =
  (\mathbf{a}_\text{p},\mathbf{b}_\text{p}, \mathbf{c}_\text{p})   \boldsymbol{P}^{-1}.
\end{align}
$(\tilde{\mathbf{a}}_\text{s},  \tilde{\mathbf{b}}_\text{s},
  \tilde{\mathbf{c}}_\text{s}) $ and $(\tilde{\mathbf{a}}_\text{p},
  \tilde{\mathbf{b}}_\text{p},\tilde{\mathbf{c}}_\text{p}) $ define a new
supercell and a new primitive cell, respectively. Because
$\det(\boldsymbol{P})=\det(\boldsymbol{Q})=1$, they have the same volumes as the
former ones, and generate the same lattices. The benefit of using those new
primitive and supercell comes from the fact that $\boldsymbol{D}$ is diagonal.
The new primitive and supercell lattice vectors are therefore collinear.

We can easily find the lattice vectors located within the new supercell. If we
write $  \boldsymbol{D} =\text{diag}(n_1, n_2, n_3)$, they are given by the
vectors
\begin{align}
  \mathbf{R}_m= (\tilde{\mathbf{a}}_\text{p}, \tilde{\mathbf{b}}_\text{p},\tilde{\mathbf{c}}_\text{p}) \begin{pmatrix} m_1\\m_2 \\ m_3\end{pmatrix} =(\tilde{\mathbf{a}}_\text{p}, \tilde{\mathbf{b}}_\text{p},\tilde{\mathbf{c}}_\text{p})  \boldsymbol{m},
\end{align}
with $m_i \in \{0, 1, \ldots, n_i - 1\}$. This can also be written
\begin{align}
  \mathbf{R}_m & =   (\tilde{\mathbf{a}}_\text{s},
  \tilde{\mathbf{b}}_\text{s}, \tilde{\mathbf{c}}_\text{s})  \boldsymbol{D}^{-1}\boldsymbol{m}                                            \\
               & =   (\mathbf{a}_\text{s},  \mathbf{b}_\text{s}, \mathbf{c}_\text{s})   \boldsymbol{Q} \boldsymbol{D}^{-1}\boldsymbol{m}.
\end{align}
Therefore, the lattice points within the original supercell have coordinates
\begin{align}
  \label{eq:supercell-lattice-points}
  \boldsymbol{x}_{s} & = \boldsymbol{Q} \boldsymbol{D}^{-1} \boldsymbol{m} \pmod
  {\mathbf{1}}
\end{align}
along the supercell lattice vectors $(\mathbf{a}_\text{s}, \mathbf{b}_\text{s},
  \mathbf{c}_\text{s})$. The $\pmod {\mathbf{1}}$ operation is used to shift the
lattice vectors from within the new supercell to within the original supercell.

In Fig.~\ref{fig:supercell-construction}, an example of a supercell
construction for a two dimensional lattice is presented. $\boldsymbol{D} =
  \boldsymbol{P} \boldsymbol{M_{\text{p}\rightarrow\text{s}}} \boldsymbol{Q}$ is
computed as
\begin{align}
  \begin{pmatrix}
    2 & 0 \\
    0 & 4
  \end{pmatrix} & =
  \begin{pmatrix}
    0  & 1 \\
    -1 & 1
  \end{pmatrix}
  \begin{pmatrix}
    2  & 2 \\
    -2 & 2
  \end{pmatrix}
  \begin{pmatrix}
    0 & -1 \\
    1 & -1
  \end{pmatrix}.
\end{align}
This gives
\begin{align}
  (\tilde{\mathbf{a}}_\text{p},
  \tilde{\mathbf{b}}_\text{p}) & = ({\mathbf{a}}_\text{p},
  {\mathbf{b}}_\text{p}) \boldsymbol{P}^{-1} =
  ({\mathbf{a}}_\text{p} +  {\mathbf{b}}_\text{p},
  -{\mathbf{a}}_\text{p}),                                 \\
  (\tilde{\mathbf{a}}_\text{s},
  \tilde{\mathbf{b}}_\text{s}) & = ({\mathbf{a}}_\text{s},
  {\mathbf{b}}_\text{s}) \boldsymbol{Q} =
  ({\mathbf{b}}_\text{s}, -{\mathbf{a}}_\text{s} -
  {\mathbf{b}}_\text{s}).
\end{align}
As can be seen in Fig.~\ref{fig:supercell-construction},
$(\tilde{\mathbf{a}}_\text{p}, \tilde{\mathbf{b}}_\text{p})$ and
$(\tilde{\mathbf{a}}_\text{s}, \tilde{\mathbf{b}}_\text{s})$ are mutually
parallel, and the new supercell is simply built by $2 \times 4$ units of the new
primitive cell. Since the lattices generated by the original or new basis
vectors are the same,  the lattice points inside the new supercell are simply
brought into the original supercell by supercell lattice translations.

Another way of constructing the supercell is to employ an auxiliary
supercell,
\begin{align}
  (\mathbf{a}_\text{aux},
  \mathbf{b}_\text{aux}, \mathbf{c}_\text{aux}) =
  (\mathbf{a}_\text{p},
  \mathbf{b}_\text{p}, \mathbf{c}_\text{p})
  \boldsymbol{M}_{\text{p}\rightarrow\text{aux}},
\end{align}
which is defined by a diagonal integer matrix
$\boldsymbol{M}_{\text{p}\rightarrow\text{aux}} = \text{diag}(n_1, n_2, n_3)$.
This gives
\begin{align}
  (\mathbf{a}_\text{aux},
  \mathbf{b}_\text{aux}, \mathbf{c}_\text{aux}) =
  (\mathbf{a}_\text{s},
  \mathbf{b}_\text{s}, \mathbf{c}_\text{s})
  \boldsymbol{M}_{\text{p}\rightarrow\text{s}}^{-1}
  \boldsymbol{M}_{\text{p}\rightarrow\text{aux}}.
\end{align}
We choose $\boldsymbol{M}_{\text{p}\rightarrow\text{s}}^{-1}
  \boldsymbol{M}_{\text{p}\rightarrow\text{aux}}$ to be an integer matrix.
To satisfy it, in the implementation, the diagonal elements
of $\boldsymbol{M}_{\text{p}\rightarrow\text{aux}}$ are determined by making
the smallest parallelepiped of $(\mathbf{a}_\text{aux}, \mathbf{b}_\text{aux},
  \mathbf{c}_\text{aux})$ that includes the parallelepiped of
$(\mathbf{a}_\text{s}, \mathbf{b}_\text{s}, \mathbf{c}_\text{s})$. The
lattice points within this auxiliary supercell are then
\begin{align}
  \mathbf{R}_m=  (\mathbf{a}_\text{p},\mathbf{b}_\text{p}, \mathbf{c}_\text{p})
  \begin{pmatrix} m_1\\m_2 \\ m_3\end{pmatrix} =
  (\mathbf{a}_\text{p},\mathbf{b}_\text{p}, \mathbf{c}_\text{p}) \boldsymbol{m},
\end{align}
with $m_i \in \{0, 1, \ldots, n_i - 1\}$. Their coordinates along the basis
vectors of the auxiliary supercell are
\begin{align}
  \boldsymbol{x}_\text{aux} =
  \boldsymbol{M}_{\text{p}\rightarrow\text{aux}}^{-1}\boldsymbol{m}
\end{align}
while their coordinates along the basis vectors of the original supercell are
\begin{align}
  \label{eq:atom-reduction-from-aux-to-p}
  \boldsymbol{x}_\text{s} = \boldsymbol{M}_{\text{p}\rightarrow\text{s}}^{-1}
  \boldsymbol{M}_{\text{p}\rightarrow\text{aux}} \boldsymbol{x}_\text{aux}
  \pmod {\mathbf{1}}.
\end{align}
As before, we used the$\pmod {\mathbf{1}}$ operation to shift the lattice points
from within the auxiliary supercell to within the original supercell. Notice
that every lattice point within the original supercell is obtained from
$\det(\boldsymbol{M}_{\text{p}\rightarrow\text{s}}^{-1}
  \boldsymbol{M}_{\text{p}\rightarrow\text{aux}})$ lattice points within the
auxiliary supercell. Only one of them should be conserved.

An example is shown in Fig.~\ref{fig:supercell-construction}. Lattice points in
the auxiliary supercell are brought into the supercell by the supercell lattice
translations. But since $\det(\boldsymbol{M}_{\text{p}\rightarrow\text{s}}^{-1}
  \boldsymbol{M}_{\text{p}\rightarrow\text{aux}})=2$, each lattice point within
the original supercell is obtained from two lattice points in the auxiliary
supercell.

For backward compatibility, to
preserve indices of atoms in the supercell, it is the latter way of the
supercell construction which is used as default in the phonopy and phono3py
codes. However, the latter algorithm is much slower than the former, and
therefore the former should be used for very larger supercells.

\subsection{Commensurate points}
From Eqs.~(\ref{eq:primitive-to-supercell}) and (\ref{eq:rcell}), we have
\begin{align}
  \label{eq:reciprocal-primitive-to-supercell}
  (\mathbf{a}_\text{p}^*,
  \mathbf{b}_\text{p}^*, \mathbf{c}_\text{p}^*) =
  (\mathbf{a}_\text{s}^*,
  \mathbf{b}_\text{s}^*, \mathbf{c}_\text{s}^*)
  \boldsymbol{M}_{\text{p}\rightarrow\text{s}}^{\intercal}.
\end{align}
$(\mathbf{a}_\text{s}^*,\mathbf{b}_\text{s}^*, \mathbf{c}_\text{s}^*)$ are the
reciprocal basis vectors associated to the supercell.  $\mathbf{q}$ points given
by integer linear combinations of the reciprocal supercell basis vectors are
called commensurate  with the supercell because they fulfill
$e^{i\mathbf{q}\cdot \mathbf{R}_L}=1$ for any supercell lattice vector $
  \mathbf{R}_L$.

For practical purpose, we are interested in the commensurate points located
within the reciprocal primitive cell. Since
Eq.~(\ref{eq:reciprocal-primitive-to-supercell}) has the same form as
Eq.~(\ref{eq:primitive-to-supercell}), the same approaches as those explained in
Sec.~\ref{seq:supercell-construction} can be used for generation of the
commensurate points within the reciprocal primitive cell.

\section{Transformation between force constants and dynamical matrices}
\label{sec:fc-to-dynmat-dynmat-to-fc}
In Eqs.~(\ref{eq:dynamical-matrix}) and (\ref{eq:phonon-eigenvalue-problem}),
force constants are transformed to phonon eigenvalues and eigenvectors. In
principle, in Eq.~(\ref{eq:dynamical-matrix}) the summation over $l'$  is
performed for all lattice vectors. However, in the supercell approach, it is not
$ \Phi_{0{\kappa}\alpha,l'{\kappa}'\alpha'}$ which is obtained from numerical
simulations, but $\Phi^\text{SC}_{0{\kappa}\alpha,l'{\kappa}'\alpha'}=\sum_{L'}
  \Phi_{0{\kappa}\alpha,L'+l'{\kappa}'\alpha'}$,\cite{Parlinski-phonon-1997} with
$l'$ restricted to the supercell. It means that the force constants are only
known within a restricted region of the crystal, the supercell, and within this
region they are contaminated by the periodic repetitions of the supercell.  This
also means that the supercell force constants
$\Phi^\text{SC}_{0{\kappa}\alpha,l'{\kappa}'\alpha'}$ are periodic under a supercell
lattice translation while the phase factor
$e^{i\mathbf{q}\cdot(\mathbf{R}^0_{l'\kappa'}-\mathbf{R}^0_{0\kappa})}$ is not,
except when $\mathbf{q}$ is a wave vector commensurate with the supercell. There
is therefore an ambiguity about where to choose the atom $(l'\kappa')$. For
commensurate wave vectors, it could be chosen in any periodic repetition of the
supercell containing atom $(0 \kappa)$ without changing the value of phase
factor. However, for other values of $\mathbf{q}$, this value would be changed,
and therefore this choice matters.

In the phonopy and phono3py codes, following
Ref.~\onlinecite{Parlinski-phonon-1997}, the phase factors given by the shortest
vectors of $\mathbf{R}^0_{l'\kappa'} + \mathbf{R}_L -\mathbf{R}^0_{0\kappa}$ are
chosen, where $\mathbf{R}_L$ are the positions of the supercell lattice points
$L$. This is implemented by searching $L$ by
\begin{align}
  \label{eq:equi-distant-pairs}
  \{L\}_{\kappa\kappa'l'}=\argmin_L (|\mathbf{R}^0_{l'\kappa'} +
  \mathbf{R}_L - \mathbf{R}^0_{0\kappa}|),
\end{align}
and the results are stored and used many times in the calculation. Since
multiple $L$ can be found for each pair of atoms ($0\kappa$) in the primitive cell
and ($l'\kappa'$)  in the supercell cell,
Eq.~(\ref{eq:equi-distant-pairs}) gives a set of $L$. The dynamical matrix of
the supercell is computed from the supercell force constants
$\Phi^\text{SC}_{0{\kappa}\alpha,l'{\kappa}'\alpha'}$  averaging the phase
factors of $\{L\}$ as follows,\cite{Parlinski-phonon-1997}
\begin{align}
  \label{eq:dynamical-matrix-from-supercell-fc}
  D^\text{SC}_{{\kappa}\alpha,{\kappa}'\alpha'}(\mathbf{q}) & =
  \frac{1}{\sqrt{m_{\kappa}m_{{\kappa}'}}} \sum_{l'}
  \Phi^\text{SC}_{0{\kappa}\alpha,l'{\kappa}'\alpha'}\nonumber
  \\
                                                            &
  \times \frac{1}{|\{L\}_{\kappa\kappa'l'}|} \sum_{\{L\}_{\kappa\kappa'l'}}
  e^{i\mathbf{q}\cdot
      (\mathbf{R}^0_{l'\kappa'}+\mathbf{R}_{L}-\mathbf{R}^0_{0\kappa})}.
\end{align}

Eq.~(\ref{eq:dynamical-matrix}) can be inverted to compute force constants from
$N$ dynamical matrices,\cite{Thermodynamics-of-crystals}
\begin{align}
  \label{eq:force-constants-from-dynmat}
  \Phi_{0\kappa\alpha,l'\kappa'\alpha'}
  = \frac{\sqrt{m_\kappa m_{\kappa'}}}{N}
  \sum_\mathbf{q} D_{\kappa\alpha,\kappa'\alpha'}(\mathbf{q})
  e^{-i\mathbf{q}\cdot(\mathbf{R}^0_{l'\kappa'}-\mathbf{R}^0_{0\kappa})}.
\end{align}
$D_{\kappa\alpha,\kappa'\alpha'}(\mathbf{q})$ can be obtained from known
eigenvectors and eigenvalues, as in Eq.~(\ref{eq:dynamical-matrix-in-matrix}),
or calculated by Eq.~(\ref{eq:dynamical-matrix-from-supercell-fc}).
$\Phi^\text{SC}_{0\kappa\alpha,l'\kappa'\alpha'}$ can be obtained from $D^\text{SC}_{\kappa\alpha,\kappa'\alpha'}(\mathbf{q})$
by restricting the above $\mathbf{q}$ sum to the wavevectors $\mathbf{q}^{\star}$ commensurate with the supercell.
Eq.~(\ref{eq:force-constants-from-dynmat}) can also be used to transform
supercell force constants in a different supercell shape by oversampling
$\mathbf{q}^\star$ points. Indeed, we obtain
\begin{widetext}
  \begin{align}
    \label{eq:supercell-force-constants-oversampling}
    \Phi^{\text{SC}_{\text{L} \leftarrow \text{S}}}_{0\kappa\alpha,l_\text{L}'\kappa'\alpha'}
    = & \frac{1}{N} \sum_{\mathbf{q}^\star_\text{L}}
    e^{-i\mathbf{q}^\star_\text{L}\cdot ( \mathbf{R}^0_{l_\text{L}'\kappa'} -  \mathbf{R}^0_{0\kappa}    )}
    \left[
    \sum_{l_\text{S}'}
    \Phi^{\text{SC}_\text{S}}_{0{\kappa}\alpha,l_\text{S}'{\kappa}'\alpha'} \frac{1}{|\{L_\text{S}\}_{\kappa\kappa'l_\text{S}'}|}
    \sum_{\{L_\text{S}\}_{\kappa\kappa'l_\text{S}'}}
    e^{i\mathbf{q}^\star_\text{L} \cdot
        (\mathbf{R}^0_{l_\text{S}'\kappa'} + \mathbf{R}_{L_\text{S}} -
        \mathbf{R}^0_{0\kappa})}
    \right],
  \end{align}
\end{widetext}
where subscripts of variables $X_\text{L}$ and $X_\text{S}$ mean those defined
in different supercell shapes, e.g., larger (L) and smaller (S) supercells,
respectively. The original supercell force constants
$\Phi^{\text{SC}_\text{S}}_{0{\kappa}\alpha,l_\text{S}'{\kappa}'\alpha'}$ are
given in the smaller supercell. Commensurate points $\mathbf{q}_\text{L}^\star$
are sampled with respect to the larger supercell. The transformed supercell
force constants $\Phi^{\text{SC}_{\text{L} \leftarrow
  \text{S}}}_{0\kappa\alpha,l_\text{L}'\kappa'\alpha'}$ are given in the larger
supercell. A possible application is to embed anharmonic contribution obtained
through self-consistent harmonic approximation using a smaller supercell into
the harmonic force constants of a larger
supercell.\cite{IXS-KCl-NaCl}

\section{Non-analytical term correction}
\label{seq:non-analytical-term-correction}

Long range dipole-dipole interactions are difficult to capture in a supercell
approach. Therefore it is treated with the help of a model, the so called
non-analytical term correction (NAC).\cite{Pick-1970, Giannozzi-DFPT-1991,
  Gonze-1994, Gonze-1997}

At the commensurate points, this contribution is already included in
Eq.~(\ref{eq:dynamical-matrix-from-supercell-fc}) via the supercell force
constants. However, at general  $\mathbf{q}$ points it is not. Gonze and
Lee~\cite{Gonze-1997} have formulated this contribution to the dynamical
matrix, and in the phonopy and phono3py codes only  the reciprocal space term
of the dipole-dipole interaction contribution to the dynamical matrix is
calculated. It is given by
\begin{align}
  \label{eq:dynmat-dipole-dipole}
  D^\text{DD}_{\kappa\alpha, \kappa'\alpha'} &
  (\mathbf{q}) = \frac{4\pi}{V_\text{c}} \frac{e^2}{4 \pi \epsilon_0}
  \frac{1}{\sqrt{m_{\kappa}m_{{\kappa}'}}} \sum_{\beta\beta'}
  Z^*_{\kappa,\beta\alpha} Z^*_{\kappa',\beta'\alpha'}
  \nonumber                                                                                                                    \\
                                             & \times
  \sum_{\mathbf{Q=G+q}}
  \frac{Q_{\beta} Q_{\beta'}}
  {\sum_{\gamma\gamma'} Q_\gamma \epsilon_{\gamma\gamma'}
    Q_{\gamma'}}
  \nonumber                                                                                                                    \\
                                             & \times e^{i\mathbf{G} \cdot (\mathbf{R}^0_{0\kappa} - \mathbf{R}^0_{0\kappa'})}
  \exp\left( -\sum_{\gamma\gamma'} \frac{
    Q_\gamma \epsilon_{\gamma\gamma'} Q_{\gamma'}} {4\Lambda^2} \right),
\end{align}
where $\beta$ and $\gamma$ indicate the Cartesian coordinates. If the
polarization is called $\mathbf{P}$, then
$Z^*_{\kappa,\beta\alpha}=\frac{V_\text{c}}{e}\frac{\partial
    \mathbf{P}_{\beta}}{\partial \mathbf{R}_{0\kappa \alpha}}|_{\mathbf{E}=0}$ are
the Born effective charges at zero electric field, $\mathbf{E}=0$.
$\epsilon_{\gamma\gamma'}$ is the high-frequency static dielectric constant
tensor. $\Lambda$ is a parameter adjusted together with the cutoff radius used
for the summation over $\mathbf{G}$. A translational invariance condition can be
applied to Eq.~(\ref{eq:dynmat-dipole-dipole}) by~\cite{Gonze-1997}

\begin{align}
  D^{\text{DD}}_{\kappa\alpha,
  \kappa'\alpha'}(\mathbf{q}) \leftarrow & D^{\text{DD}}_{
      \kappa\alpha, \kappa'\alpha'}(\mathbf{q})
  - \delta_{\kappa\kappa'} \sum_{\kappa''} \sqrt{\frac{m_{\kappa''}}{m_{\kappa}} }
  D^{\text{DD}}_{\kappa\alpha,
      \kappa''\alpha'}(\mathbf{q=0}).
\end{align}

As said already, at the commensurate points, this dipole-dipole contribution is
already included in Eq.~(\ref{eq:dynamical-matrix-from-supercell-fc}) via the
supercell force constants. Therefore, Eq.~(\ref{eq:dynmat-dipole-dipole}) is
used for the interpolation of the dynamical matrix at general
$\mathbf{q}$ points. The procedure is reported in Ref.~\onlinecite{Gonze-1997},
which is described shortly as follows. At the commensurate points
$\mathbf{q}^\star$, the short-range dynamical matrix is calculated as
\begin{align}
  \label{eq:short-range-dynmat}
  D^{\text{SR,SC}}_{{\kappa}\alpha,{\kappa}'\alpha'}(\mathbf{q}^\star) =
  D^\text{SC}_{{\kappa}\alpha,{\kappa}'\alpha'}(\mathbf{q}^\star)
  - D^\text{DD}_{\kappa\alpha, \kappa'\alpha'} (\mathbf{q}^\star).
\end{align}
Next, the short-range supercell force constants
$\Phi^{\text{SR,SC}}_{0\kappa\alpha,l'\kappa'\alpha'}$ is obtained from
$\{D^{\text{SR,SC}}_{{\kappa}\alpha,{\kappa}'\alpha'}
  (\mathbf{q}^\star)\}$ using
Eq.~(\ref{eq:force-constants-from-dynmat}), then the short-range
dynamical matrix
$D^{\text{SR,SC}}_{{\kappa}\alpha,{\kappa}'\alpha'}(\mathbf{q})$ is calculated
at the general $\mathbf{q}$ point by
Eq.~(\ref{eq:dynamical-matrix-from-supercell-fc}), i.e.,
\begin{align}
  \{D^{\text{SR,SC}}_{{\kappa}\alpha,{\kappa}'\alpha'}(\mathbf{q}^\star)\}
  \xrightarrow[\text{Eq.(\ref{eq:force-constants-from-dynmat})}]{}
  \Phi^{\text{SR,SC}}_{0\kappa\alpha,l'\kappa'\alpha'}
  \xrightarrow[\text{Eq.(\ref{eq:dynamical-matrix-from-supercell-fc})}]{}
  D^{\text{SR,SC}}_{{\kappa}\alpha,{\kappa}'\alpha'}(\mathbf{q}).
\end{align}
Finally, the dynamical matrix at the general $\mathbf{q}$ point with NAC is
obtained by
\begin{align}
  \label{eq:long-range-dynmat}
  D^\text{NAC}_{{\kappa}\alpha,{\kappa}'\alpha'}(\mathbf{q}) =
  D^{\text{SR,SC}}_{{\kappa}\alpha,{\kappa}'\alpha'}(\mathbf{q})
  + D^\text{DD}_{\kappa\alpha, \kappa'\alpha'} (\mathbf{q}).
\end{align}

\begin{figure}[ht]
  \begin{center}
    \includegraphics[width=0.8\linewidth]{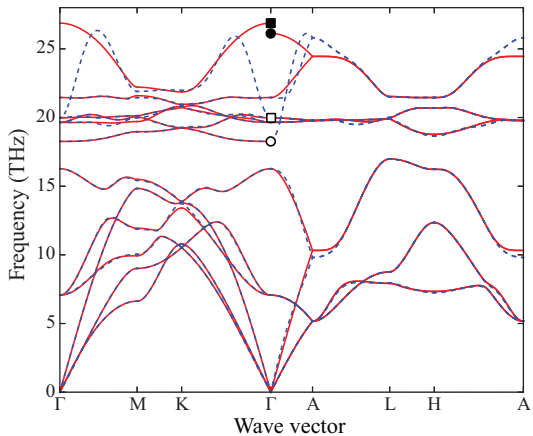}
    \caption{Calculated phonon band structure of wurtzite-type AlN using the
      $5\times 5\times 3$ supercell with (solid curves) and without (dashed
      curves) non-analytical term correction (NAC). Circle and square symbols
      depict phonon frequencies at $\mathbf{q} \rightarrow \mathbf{0}$, where
      filled and open symbols indicate phonon modes calculated with and without
      NAC, respectively. Wave vector path was selected by
      Seek-path.\cite{seekpath,spglib}
      \label{fig:AlN-band}
    }
  \end{center}
\end{figure}

An example of application of NAC to wurtzite-type AlN is shown in
Fig.~\ref{fig:AlN-band}. As can be seen, the correction is significant near
the $\Gamma$ point. It also shows the directional dependence near the
$\Gamma$, in the direction of K and A, due to the non-spherical symmetry of
$\epsilon_{\gamma\gamma'}$ and $Z^*_{\kappa,\beta\alpha}$ of AlN.

\section{Regular grid in reciprocal primitive cell}
\label{sec:regular-grids}
\begin{figure}[ht]
  \begin{center}
    \includegraphics[width=1.00\linewidth]{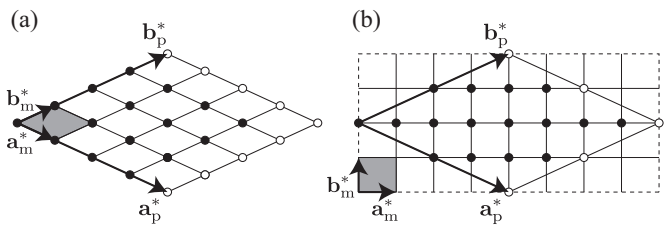}
    \caption{Different $\Gamma$-centred regular grids in plane reciprocal
      primitive cell ($c2mm$). Filled circle symbols depict the grid points. Open
      circle symbols show the grid points equivalent to some other grid points by
      periodicity. Shaded area indicates a
      microzone~\cite{MacDonald-tetrahedron-1979} and $\mathbf{a}_\text{m}^*$ and
      $\mathbf{b}_\text{m}^*$ give the basis vectors of the microzone.  (a)
      Traditional regular grid. (b) A type of generalized regular
      grid.\cite{Moreno-grg-1992, Wisesa-grg-2016, Morgan-grg-2018,
        Morgan-grg-2020} \label{fig:regular-grid} }
  \end{center}
\end{figure}

When phonon properties at $\mathbf{q}$ points need to be integrated over the BZ,
a technique often used is to discretize the reciprocal space. Usually reciprocal
primitive cells are uniformly sampled by a regular grid such as
Ref.~\onlinecite{Monkhorst-Pack}, even if other choices are
possible.\cite{Chadi-Cohen-1973} Traditionally, the regular grid is defined by
evenly dividing the reciprocal basis vectors, as shown in
Fig.~\ref{fig:regular-grid} (a). However, other regular grids may be chosen, as
shown in Fig.~\ref{fig:regular-grid} (b). For example, the grid shown in
Fig.~\ref{fig:regular-grid} (b) is defined by dividing evenly the reciprocal
basis vectors of the conventional unit cell instead. This is a type of
generalized regular grid,\cite{Moreno-grg-1992, Wisesa-grg-2016, Morgan-grg-2018,
  Morgan-grg-2020} as will be explained later.

\subsection{Traditional regular grid}
\label{sec:traditional-regular-grid}
Traditionally, the volume of the reciprocal primitive cell is divided into
uniform microzones so that the basis vectors of the reciprocal primitive cell
are simply integer multiples of the basis vectors of each microzone
$(\mathbf{a}_\text{m}^*, \mathbf{b}_\text{m}^*, \mathbf{c}_\text{m}^*)$. The
equation which defines the microzone basis vectors is then

\begin{align}
  \label{eq:simple-microzone}
  (\mathbf{a}_\text{p}^*, \mathbf{b}_\text{p}^*, \mathbf{c}_\text{p}^*) =
  (\mathbf{a}_\text{m}^*, \mathbf{b}_\text{m}^*, \mathbf{c}_\text{m}^*) \boldsymbol{D}
\end{align}
with $\boldsymbol{D}= \text{diag}(n_1, n_2, n_3), \; n_i \in \mathbb{N}$.

A 2D example of this microzone is shown in Fig.~\ref{fig:regular-grid} (a). The
$\mathbf{q}$ points of the grid points are represented by integer linear
combinations of the microzone basis vectors plus a possible rigid shift $(s_1,
  s_2, s_3)^\intercal$. Therefore their coordinates in the
$(\mathbf{a}_\text{p}^*, \mathbf{b}_\text{p}^*, \mathbf{c}_\text{p}^*)$ basis
are
\begin{align}
  \label{eq:grid-generation}
  \begin{pmatrix}
    q_1 \\
    q_2 \\
    q_3
  \end{pmatrix}
   & = \begin{pmatrix}
         (m_1 + s_1) / n_1 \\
         (m_2 + s_2) / n_2 \\
         (m_3 + s_3) / n_3
       \end{pmatrix},
  \\ \nonumber
   & m_i \in \{0, 1, \ldots, n_i - 1\}, \; 0 \leq s_i < 1.
\end{align}
To conserve symmetry, $n_i$ and $s_i$ are chosen so that the microzone lattice
with the shift is invariant under the crystallographic point group $\mathbb{P}$.

\subsection{Generalized regular grid}
\label{seq:generalized-regular-grid}
A generalized regular grid is defined using a conventional unit cell related to
the primitive cell by Eq.~(\ref{eq:conv-to-prim-matrix}). The reciprocal
conventional unit cell is therefore
\begin{align}
  \label{eq:relation-reciprocal-cells}
  (\mathbf{a}_\text{c}^*, \mathbf{b}_\text{c}^*, \mathbf{c}_\text{c}^*) =
  (\mathbf{a}_\text{p}^*, \mathbf{b}_\text{p}^*, \mathbf{c}_\text{p}^*)
  \boldsymbol{P}_\text{prim}^{\intercal}.
\end{align}
Microzones can be defined considering that their basis vectors are integer
divisions of the reciprocal basis vectors of the conventional unit cell. This is
written as
\begin{align}
  \label{eq:generaized-microzone}
  (\mathbf{a}_\text{c}^*, \mathbf{b}_\text{c}^*, \mathbf{c}_\text{c}^*) =
  (\mathbf{a}_\text{m}^*, \mathbf{b}_\text{m}^*, \mathbf{c}_\text{m}^*)
  \text{diag}(
  N_1,
  N_2,
  N_3),\; N_i \in \mathbb{N}.
\end{align}
A 2D example of this microzone is shown in Fig.~\ref{fig:regular-grid} (b). From
Eqs.~(\ref{eq:relation-reciprocal-cells}) and (\ref{eq:generaized-microzone}),
we have
\begin{align}
  \label{eq:grid-matrix}
  (\mathbf{a}_\text{p}^*, \mathbf{b}_\text{p}^*, \mathbf{c}_\text{p}^*) & =
  (\mathbf{a}_\text{m}^*, \mathbf{b}_\text{m}^*, \mathbf{c}_\text{m}^*)
  \text{diag}(
  N_1,
  N_2,
  N_3)
  \boldsymbol{P}_\text{prim}^{-\intercal}
  \nonumber                                                                 \\
                                                                        & =
  (\mathbf{a}_\text{m}^*, \mathbf{b}_\text{m}^*, \mathbf{c}_\text{m}^*)
  \boldsymbol{M_\text{g}}.
\end{align}
The grid matrix $\boldsymbol{M_\text{g}}=\text{diag}(N_1, N_2, N_3)
  \boldsymbol{P}_\text{prim}^{-\intercal}$ is an integer matrix.
$\det(\boldsymbol{M_\text{g}})$ is the number of translationally nonequivalent
grid points in the reciprocal primitive cell. When $\boldsymbol{M_\text{g}}$ is
not a diagonal matrix, the grid generated by Eq.~(\ref{eq:grid-matrix}) is a
generalized regular grid, otherwise we obtain the traditional regular grid as
described in Sec.~\ref{sec:traditional-regular-grid}. Notice however that the
generalized regular grid may be transformed into the traditional regular grid of
some reciprocal cell using an SNF kind of decomposition, as explained in
Sec.~\ref{seq:supercell-construction}. This is shown in the next section.

\subsection{Indexing of grid points}
\label{sec:grid-indexing}
The integer matrix $\boldsymbol{M_\text{g}}$ can be reduced to a diagonal
integer matrix such as $\boldsymbol{D} = \boldsymbol{P} \boldsymbol{M_\text{g}}
  \boldsymbol{Q}$ as has been employed in Sec.~\ref{seq:supercell-construction}.
This property of the matrix decomposition is similarly used to index grid points
by integer numbers of $\{0, 1, \ldots, \det(\boldsymbol{D}) -
  1\}$~\cite{Gus-SNF-2008} in the phono3py code. Note that $\det(\boldsymbol{D}) =
  \det(\boldsymbol{M_\text{g}})$. Eq.~(\ref{eq:grid-matrix}) is rewritten as
\begin{align}
  \label{eq:oblique-lattice}
   & (\mathbf{a}_\text{p}^*, \mathbf{b}_\text{p}^*, \mathbf{c}_\text{p}^*)
  \boldsymbol{Q} =
  (\mathbf{a}_\text{m}^*, \mathbf{b}_\text{m}^*, \mathbf{c}_\text{m}^*)
  \boldsymbol{P}^{-1} \boldsymbol{D}.
\end{align}
Denoting
\begin{align}
  \label{eq:rec-basis-vectors-Q}
  (\tilde{\mathbf{a}}_\text{p}^*, \tilde{\mathbf{b}}_\text{p}^*,
  \tilde{\mathbf{c}}_\text{p}^*)
   & =(\mathbf{a}_\text{p}^*, \mathbf{b}_\text{p}^*, \mathbf{c}_\text{p}^*)
  \boldsymbol{Q},                                                           \\
  \label{eq:rec-basis-vectors-P}
  (\tilde{\mathbf{a}}_\text{m}^*, \tilde{\mathbf{b}}_\text{m}^*,
  \tilde{\mathbf{c}}_\text{m}^*)
   & =(\mathbf{a}_\text{m}^*, \mathbf{b}_\text{m}^*, \mathbf{c}_\text{m}^*)
  \boldsymbol{P}^{-1},
\end{align}
we have
\begin{align}
  \label{eq:rec-basis-vectors-oblique}
  (\tilde{\mathbf{a}}_\text{p}^*, \tilde{\mathbf{b}}_\text{p}^*,
  \tilde{\mathbf{c}}_\text{p}^*) =
  (\tilde{\mathbf{a}}_\text{m}^*, \tilde{\mathbf{b}}_\text{m}^*,
  \tilde{\mathbf{c}}_\text{m}^*) \boldsymbol{D}.
\end{align}
Since $\boldsymbol{Q}$ is a unimodular matrix,
$(\tilde{\mathbf{a}}_\text{p}^*, \tilde{\mathbf{b}}_\text{p}^*,
  \tilde{\mathbf{c}}_\text{p}^*)$ and $(\mathbf{a}_\text{p}^*,
  \mathbf{b}_\text{p}^*, \mathbf{c}_\text{p}^*)$ generate the same reciprocal
primitive lattice. Similarly $(\tilde{\mathbf{a}}_\text{m}^*,
  \tilde{\mathbf{b}}_\text{m}^*, \tilde{\mathbf{c}}_\text{m}^*)$ and
$(\mathbf{a}_\text{m}^*, \mathbf{b}_\text{m}^*, \mathbf{c}_\text{m}^*)$
generate the same microzone lattice due to the unimodular matrix
$\boldsymbol{P}^{-1}$.

Eq.~(\ref{eq:rec-basis-vectors-oblique}) has the same form as
Eq.~(\ref{eq:simple-microzone}) with $\boldsymbol{D} = \text{diag}(n_1, n_2,
  n_3)$. Therefore the $\mathbf{q}$ points of the grid points are calculated
similarly as Eq.~(\ref{eq:grid-generation}) but in the basis
$(\tilde{\mathbf{a}}_\text{p}^*, \tilde{\mathbf{b}}_\text{p}^*, \tilde{\mathbf{c}}_\text{p}^*)$,

\begin{align}
  \mathbf{q}=
  (\tilde{\mathbf{a}}_\text{p}^*, \tilde{\mathbf{b}}_\text{p}^*, \tilde{\mathbf{c}}_\text{p}^*)
  \begin{pmatrix}
    (m_1 + \tilde{s}_1) / n_1 \\
    (m_2 + \tilde{s}_2) / n_2 \\
    (m_3 + \tilde{s}_3) / n_3
  \end{pmatrix}.
\end{align}
Notice that in the above equation we use $\tilde{s}_i$ rather than $s_i$. In
fact for practical purpose,  it may be convenient to define the grid shift
$\boldsymbol{s} = (s_1, s_2, s_3)^\intercal$ in the basis
$(\mathbf{a}_\text{m}^*, \mathbf{b}_\text{m}^*, \mathbf{c}_\text{m}^*)$.
Therefore we have  $\tilde{ \boldsymbol{s}}=\boldsymbol{P} \boldsymbol{s}$.

The indexing of the grid points in the reciprocal primitive cell
$(\tilde{\mathbf{a}}_\text{p}^*, \tilde{\mathbf{b}}_\text{p}^*,
  \tilde{\mathbf{c}}_\text{p}^*)$ is a trivial task. Indeed, in the phonopy and
phono3py codes, each grid point is bijectively mapped to an integer $p$ by
\begin{align}
  \label{eq:grid-indexing}
  p = m_1 + n_1 m_2 +  n_1 n_2 m_3, \;  m_i \in \{0, 1, \ldots, n_i -1\}.
\end{align}
With this defintion, $0 \leq p < n_1 n_2 n_3$.

The $\mathbf{q}$ points generated this way are located within the reciprocal
cell $  (\tilde{\mathbf{a}}_\text{p}^*, \tilde{\mathbf{b}}_\text{p}^*,
  \tilde{\mathbf{c}}_\text{p}^*) $. They could be shifted to the reciprocal cell
$(\mathbf{a}_\text{p}^*, \mathbf{b}_\text{p}^*, \mathbf{c}_\text{p}^*)$ by
reciprocal lattice vector translations if needed. Notice however that to obtain
the integer $p$ through Eq.~(\ref{eq:grid-indexing}) for a general integer
triplet $ (m_1, m_2, m_3)^\intercal$, modulo $\boldsymbol{n}=(n_1, n_2,
  n_3)^\intercal$ is required to locate the point within the cell $
  (\tilde{\mathbf{a}}_\text{p}^*, \tilde{\mathbf{b}}_\text{p}^*,
  \tilde{\mathbf{c}}_\text{p}^*)$. Indeed, $ (m_1, m_2, m_3)^\intercal$ and $(m_1
  + G_1 n_1, m_2 + G_2 n_2, m_3 + G_3 n_3)^\intercal$ indicate different locations
in $\mathbf{q}$ space, although they are equivalent points due to periodicity.

\subsection{Symmetry of generalized regular grids}
\label{sec:symmetry-of-regular-grid}
For a $\Gamma$ centred grid ($s_i=0$), the coordinates of a $\mathbf{q}$ point
in the basis $(\mathbf{a}_\text{p}^*, \mathbf{b}_\text{p}^*,
  \mathbf{c}_\text{p}^*)$ are given by
\begin{align}
  \label{eq:grg-grid-point}
  \boldsymbol{q} = \boldsymbol{Q} \boldsymbol{D}^{-1} \boldsymbol{m} +
  \boldsymbol{G},
\end{align}
where $\boldsymbol{m} = (m_1, m_2, m_3)^\intercal$ and the reciprocal lattice
vector $\boldsymbol{G}$ is chosen to bring $q_i$ in the interval $[0,1)$. If the
regular grid is a traditional one, we simply have $\boldsymbol{Q}=\boldsymbol{P}
  =\boldsymbol{1} $. As shown by Eq.~(\ref{qprime}), the image of this
$\mathbf{q}$ point through an operation of the crystallographic point group is
given by
\begin{align}
  \boldsymbol{q}'= \boldsymbol{S}^{-\intercal}\boldsymbol{Q} \boldsymbol{D}^{-1} \boldsymbol{m}+ \boldsymbol{S}^{-\intercal}\boldsymbol{G}+\boldsymbol{G}' \label{qp1}
\end{align}
where we have added a reciprocal lattice vector $\boldsymbol{G}'$ to shift
$q'_i$ of the image point in the interval $[0,1)$. If  $\boldsymbol{q}'$
belongs to the grid we have defined, for all $ \boldsymbol{S}^{-\intercal}$ in
the crystallographic point group, we will say that the grid is invariant under
the crystallographic point group. If it is so, $\boldsymbol{q}'$ can be written
as
\begin{align}
  \boldsymbol{q}' = \boldsymbol{Q} \boldsymbol{D}^{-1} \boldsymbol{m}' +   \boldsymbol{G}''. \label{qp2}
\end{align}
with $m'_i \in \{0, 1, \ldots, n_i -1\}$.
Comparing Eq.~(\ref{qp1}) and Eq.~(\ref{qp2}), we obtain
\begin{align}
  \boldsymbol{m}' & =
  (\boldsymbol{Q} \boldsymbol{D}^{-1})^{-1}
  \boldsymbol{S}^{-\intercal}(\boldsymbol{Q} \boldsymbol{D}^{-1})
  \boldsymbol{m}                                                \\
                  & + (\boldsymbol{Q} \boldsymbol{D}^{-1})^{-1}
  (\boldsymbol{S}^{-\intercal}\boldsymbol{G}+\boldsymbol{G}' -  \boldsymbol{G}'')
\end{align}
Because $\boldsymbol{Q}$ and $\boldsymbol{S}^{-\intercal}$ are unimodular, and $
  \boldsymbol{D}$ contains the number of divisions along
$(\tilde{\mathbf{a}}_\text{p}^*, \tilde{\mathbf{b}}_\text{p}^*,
  \tilde{\mathbf{c}}_\text{p}^*)$, the last term is always a reciprocal lattice
vector. The matrix
\begin{align}
  \tilde{\boldsymbol{S}}^{-\intercal} =
  (\boldsymbol{Q} \boldsymbol{D}^{-1})^{-1}
  \boldsymbol{S}^{-\intercal}(\boldsymbol{Q} \boldsymbol{D}^{-1})
  \label{S_tilde}
\end{align}
is the matrix representation of $\mathcal{S}$ in the $
  (\tilde{\mathbf{a}}_\text{m}^*, \tilde{\mathbf{b}}_\text{m}^*,
  \tilde{\mathbf{c}}_\text{m}^*)$ basis. Its determinant is always $1$ or $-1$,
but for $\boldsymbol{m}' $ to be integer, it has to have integer entries, and
therefore be unimodular. This is checked in the phono3py code, and if it is
true for all $\boldsymbol{S}^{-\intercal}$ in crystallographic point group,
the regular grid follows the crystallographic point group, and we consider it
is properly defined. The last equation can also be written as
\begin{align}
  \boldsymbol{P}^{-1}\tilde{\boldsymbol{S}}^{-\intercal} \boldsymbol{P}=
  \boldsymbol{M}_{\text{g}}\boldsymbol{S}^{-\intercal}\boldsymbol{M}_{\text{g}}^{-1}
\end{align}
and an equivalent strategy is to check that both sides are unimodular.

Introducing the subgrid shift $\boldsymbol{s}$ (see Sec.
\ref{sec:grid-indexing}) can further break the symmetry of the regular grid. In
Eq.~(\ref{eq:grg-grid-point}) $\boldsymbol{m}$ is replaced by $\boldsymbol{m} +
  \boldsymbol{P} \boldsymbol{s}$ and $\boldsymbol{m}'$ by $\boldsymbol{m}' +
  \boldsymbol{P} \boldsymbol{s}$ in Eq.~(\ref{qp2}). Requiring the shift to be
invariant $\pmod{\mathbf{1}}$ under an operation of the crystallographic point
group, $\tilde{\boldsymbol{S}}^{-\intercal} \boldsymbol{P}
  \boldsymbol{s}=\boldsymbol{P} \boldsymbol{s} \pmod{\mathbf{1}}$, we obtain the
condition
\begin{align}
  \boldsymbol{P}^{-1}\tilde{\boldsymbol{S}}^{-\intercal}
  \boldsymbol{P}\boldsymbol{s}=
  \boldsymbol{M}_{\text{g}}\boldsymbol{S}^{-\intercal}\boldsymbol{M}_{\text{g}}^{-1}
  \boldsymbol{s}=\boldsymbol{s} \pmod{\mathbf{1}}
\end{align}
against all $\boldsymbol{S}^{-\intercal}$ in the crystallographic point group.

\subsection{Double grid for subgrid shift}
\label{sec:double-grid}
To satisfy the crystallographic symmetry, the subgrid shift $s_i$ is normally
chosen to be either 0 or $1/2$. It is better to treat grid point arithmetic by
integers for the computer implementation and its performance. This is realized
by doubling $\boldsymbol{m}$ and $\boldsymbol{s}$. This well-known technique,
e.g., presented in Ref.~\onlinecite{Blochl-tetrahedron-1994}, is directly usable
for the generalized regular grid in Sec.~\ref{seq:generalized-regular-grid}. As
a choice, we define it by
\begin{align}
  \label{eq:double-grid-address}
  \boldsymbol{m}^\text{d} = 2(\boldsymbol{m} + \boldsymbol{P} \boldsymbol{s}),
  \pmod {2\boldsymbol{n}},
\end{align}
and the $\mathbf{q}$ points are given like in Eq.~(\ref{eq:grg-grid-point}) by
\begin{align}
  \boldsymbol{q} = \boldsymbol{Q} \boldsymbol{D}^{-1} \boldsymbol{m}^\text{d}
  /2 + \boldsymbol{G}.
\end{align}
The symmetry operation is implemented as
\begin{align}
  \label{eq:rotation-double-grid}
  {\boldsymbol{m}^\text{d}}' =\tilde{\boldsymbol{S}}^{-\intercal}
  \boldsymbol{m}^\text{d} \pmod{2\boldsymbol{n}}.
\end{align}
The index of the grid point ${\boldsymbol{m}^\text{d}}'$ is obtained by
Eq.~(\ref{eq:grid-indexing}) after recovering $\boldsymbol{m}'$ using
Eq.~(\ref{eq:double-grid-address}),
\begin{align}
  \label{eq:double-to-single-grid}
  \boldsymbol{m}'=({\boldsymbol{m}^\text{d}}' - 2\boldsymbol{P} \boldsymbol{s})/2
  \pmod
  {\boldsymbol{n}}.
\end{align}
In the
phono3py code, an integer vector of $2\boldsymbol{s}$ is used
for the implementation, instead of  $\boldsymbol{s}$, to avoid floating point arithmetic.

\subsection{Grid points in BZ}
\label{sec:grid-points-1st-BZ}
In this section, the BZ grid points are defined to include different $\mathbf{q}$
points on the BZ surface that may be equivalent grid points by reciprocal
lattice translations, in addition to the grid points inside the BZ.

For a $\Gamma$ centred grid, in the $  (\tilde{\mathbf{a}}_\text{m}^*,
\tilde{\mathbf{b}}_\text{m}^*, \tilde{\mathbf{c}}_\text{m}^*)$ basis,
each BZ grid point is represented by the integer triplet
\begin{align}
  \boldsymbol{m}^\text{BZ}=
  \begin{pmatrix}
    m^\text{BZ}_1 \\ m^\text{BZ}_2 \\ m^\text{BZ}_3
  \end{pmatrix} =
  \begin{pmatrix}
    m_1 + G^{\boldsymbol{m}}_{1} n_1 \\
    m_2 + G^{\boldsymbol{m}}_{2} n_2 \\
    m_3 + G^{\boldsymbol{m}}_3 n_3
  \end{pmatrix}, \;
  G^{\boldsymbol{m}}_{i} \in \mathbb{Z},
\end{align}
where $\boldsymbol{G}^{\boldsymbol{m}}$ is chosen to minimize the norm of
$\mathbf{q}=(\tilde{\mathbf{a}}_\text{p}^*,
  \tilde{\mathbf{b}}_\text{p}^*, \tilde{\mathbf{c}}_\text{p}^*)
  \boldsymbol{D}^{-1} \boldsymbol{m}^\text{BZ}=(\tilde{\mathbf{a}}_\text{m}^*,
  \tilde{\mathbf{b}}_\text{m}^*, \tilde{\mathbf{c}}_\text{m}^*)
  \boldsymbol{m}^\text{BZ}$. Therefore, $\boldsymbol{m}^\text{BZ}$ is written as
\begin{align}
  \label{eq:bz-grid-point}
  \boldsymbol{m}^\text{BZ} = \boldsymbol{m}  +
  \boldsymbol{D} \argmin_{\boldsymbol{G}^{\boldsymbol{m}}}
  (|(\tilde{\mathbf{a}}_\text{p}^*,
  \tilde{\mathbf{b}}_\text{p}^*,
  \tilde{\mathbf{c}}_\text{p}^*)(\boldsymbol{D}^{-1} \boldsymbol{m} +
  \boldsymbol{G}^{\boldsymbol{m}})|).
\end{align}
Multiple $\boldsymbol{m}^\text{BZ}$ can be found when the grid point is on the
BZ surface.

In the phono3py code, Eq.~(\ref{eq:bz-grid-point}) is implemented as
follows. As the first step, reduced basis vectors of the reciprocal primitive
cell are obtained by using the Niggli reduction or any reduction scheme. This is
written using the change-of-basis matrix
$\boldsymbol{M}^*_{\text{p}\rightarrow\text{r}}$ as
\begin{align}
  (\mathbf{a}_\text{r}^*, \mathbf{b}_\text{r}^*,
  \mathbf{c}_\text{r}^*) = (\mathbf{a}_\text{p}^*,
  \mathbf{b}_\text{p}^*, \mathbf{c}_\text{p}^*)
  \boldsymbol{M}^*_{\text{p}\rightarrow\text{r}},
\end{align}
where  $\boldsymbol{M}^*_{\text{p}\rightarrow\text{r}}$ is unimodular. This
gives
\begin{align}
  \mathbf{q} & =(\mathbf{a}_\text{r}^*, \mathbf{b}_\text{r}^*,
  \mathbf{c}_\text{r}^*) (\boldsymbol{M}^*_{\text{p}\rightarrow\text{r}})^{-1}
  \boldsymbol{Q}\boldsymbol{D}^{-1} [\boldsymbol{m}+ \boldsymbol{D}
  \boldsymbol{G}^{\boldsymbol{m}}]
  \\
             &
  =(\mathbf{a}_\text{r}^*, \mathbf{b}_\text{r}^*,
  \mathbf{c}_\text{r}^*) \left[ (\boldsymbol{M}^*_{\text{p}\rightarrow\text{r}})^{-1}
  \boldsymbol{Q}\boldsymbol{D}^{-1}\boldsymbol{m} +  \boldsymbol{G} \right].
\end{align}
The second line defines the reciprocal lattice vector $\boldsymbol{G}$.
$\boldsymbol{G}$ is divided into two pieces,
\begin{align}
  \boldsymbol{G} =
  \bar{\boldsymbol{G}} + \Delta \boldsymbol{G}.
\end{align}
Nearest
integers of $-(\boldsymbol{M}^*_{\text{p}\rightarrow\text{r}})^{-1}
  \boldsymbol{Q}\boldsymbol{D}^{-1}\boldsymbol{m}$ are stored in
$\bar{\boldsymbol{G}}$, which is formally written as
\begin{align}
  \label{eq:BZ-nint}
  \bar{\boldsymbol{G}} = - \text{nint}[
  (\boldsymbol{M}^*_{\text{p}\rightarrow\text{r}})^{-1}
  \boldsymbol{Q}\boldsymbol{D}^{-1}\boldsymbol{m}]
\end{align}
to bring the following $\bar{\mathbf{q}}$ closer to the origin,
\begin{align}
  \label{eq:BZ-q-after-nint}
  \bar{\mathbf{q}} = (\mathbf{a}_\text{r}^*, \mathbf{b}_\text{r}^*,
  \mathbf{c}_\text{r}^*) \left[
  (\boldsymbol{M}^*_{\text{p}\rightarrow\text{r}})^{-1}
  \boldsymbol{Q}\boldsymbol{D}^{-1}\boldsymbol{m} +
  \bar{\boldsymbol{G}} \right].
\end{align}
$\Delta \boldsymbol{G}$ that minimizes $|\bar{\mathbf{q}} +
  (\mathbf{a}_\text{r}^*, \mathbf{b}_\text{r}^*, \mathbf{c}_\text{r}^*) \Delta
  \boldsymbol{G}|$ is searched in $\Delta G_i \in \{-2, -1, 0, 1, 2\}$. Finally,
$\boldsymbol{m}^\text{BZ}$ is obtained as
\begin{align}
  \boldsymbol{m}^\text{BZ} = \boldsymbol{m} + \boldsymbol{D}
  \boldsymbol{Q}^{-1}
  \boldsymbol{M}^*_{\text{p}\rightarrow\text{r}} (\bar{\boldsymbol{G}} +
  \Delta  \boldsymbol{G}).
\end{align}
As written above, $\boldsymbol{m}$ and $p$ are in one-to-one correspondence and
multiple $\boldsymbol{m}^\text{BZ}$ can be found for each $\boldsymbol{m}$ or
$p$. In the phono3py code, $\{\{\boldsymbol{m}^\text{BZ}(p)\}|0 \leq p <
\det(\boldsymbol{D})\}$ is calculated and stored at the initial step.

With the subgrid shift, $\boldsymbol{m}$ in the equations above are replaced by
  $\boldsymbol{m} + \boldsymbol{P} \boldsymbol{s}$. Either with or without the
  subgrid shift, $\boldsymbol{m}^\text{BZ}$ is determined using the double grid
  technique presented in Sec.~\ref{sec:double-grid}. From stored
  $\boldsymbol{m}^\text{BZ}$, the BZ $\mathbf{q}$ points are obtained by
\begin{align}
  \mathbf{q}=(\mathbf{a}_\text{p}^*, \mathbf{b}_\text{p}^*,
  \mathbf{c}_\text{p}^*) \boldsymbol{Q}\boldsymbol{D}^{-1}(\boldsymbol{m}^\text{BZ}
  + \boldsymbol{P} \boldsymbol{s}).
\end{align}

\subsection{Irreducible grid points}
\label{sec:irredubile-grid-points}
Using symmetry properties of phonons, e.g., Eq.~(\ref{eq:eigenvector-rotation}),
the computation required for the phonon calculations can be reduced. Indeed,
many phonon properties result from the BZ integration of functions of a single
$\mathbf{q}$ variable. Collecting the values of those functions for the
$\mathbf{q}$ points in the irreducible part of the BZ only may speed up the
calculation and also save memory space. When the BZ integration is performed on
a regular grid, the irreducible $\mathbf{q}$ points are obtained from the
irreducible grid points. In this section, we explain how to obtain those
irreducible grid points, where it is assumed that the regular grid satisfies the
symmetry as described in Sec.~\ref{sec:symmetry-of-regular-grid}.

\begin{figure}[ht]
  \begin{center}
    \includegraphics[width=1.0\linewidth]{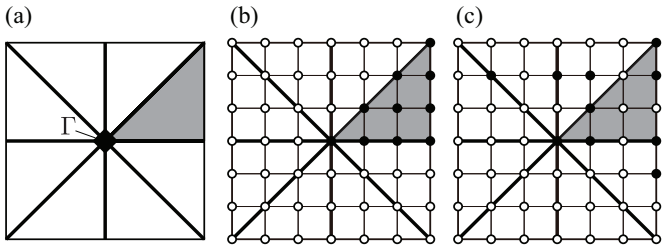}
    \caption{ (a) 2D Brillouin zone (BZ). Shaded area is a choice of the
      irreducible BZ. (b) $6 \times 6$ regular grid on the BZ including the BZ
      surface. Filled and open circle symbols show the grid points belonging to
      the irreducible BZ under the $C_\text{4v}$ symmetry and the other grid
      points, respectively. (c) The irreducible grid points may not be located
      in a connected space.
      \label{fig:grid-symmetry}
    }
  \end{center}
\end{figure}

A 2D example of BZ is presented in Fig.~\ref{fig:grid-symmetry}. A
symmetry operation acts on a $\mathbf{q}$ point as $\mathbf{q}' =
  \mathcal{S}\mathbf{q}$. The star of $\mathbf{q}$ is the set of those points
written as $\{\mathcal{S}\mathbf{q}|\mathcal{S} \in \mathbb{P}\}$, where one
$\mathbf{q}$ point represents its star. The irreducible BZ is the set of
representative $\mathbf{q}$ points of all stars in the BZ. In
Fig.~\ref{fig:grid-symmetry} (a), the 2D BZ has the eight-fold symmetry of
$C_\text{4v}$, and the irreducible BZ is depicted by the shaded area.

In a phonon calculation, $\mathbf{q}$ points are sampled on a regular grid, and
the irreducible grid points are defined as the grid points that belong to the
irreducible BZ as shown in Fig.~\ref{fig:grid-symmetry} (b). In practical
calculations, the irreducible grid points considered may however not be located
in a connected space, as shown in Fig.~\ref{fig:grid-symmetry} (c). This simply
means that the selected representative point in the star is located somewhere
else.

In the phonopy and phono3py codes, the irreducible grid points are obtained as
follows. Each grid point indexed by $p$ is represented in the double grid of
Eq.~(\ref{eq:double-grid-address}). All symmetry operations
$\tilde{\boldsymbol{S}}^{-\intercal}$ of the crystallographic point group are
applied to the grid point as given by Eq.~(\ref{eq:rotation-double-grid}). The
grid point index of the rotated grid point, $p(\mathcal{S})$, is recovered
successively applying Eqs.~(\ref{eq:double-to-single-grid}) and
(\ref{eq:grid-indexing}). The minimum value in $\{p(\mathcal{S}) |
  p(\mathcal{S}) \leq p, \mathcal{S} \in \mathbb{P} \}$ is chosen as the
irreducible grid point. The mappings of the grid point indices $p
  \xrightarrow[]{\{\mathcal{S}\}} \underline{p}$ are stored for all grid points in
$0 \leq p < \det(\boldsymbol{D})$, and the unique elements of
$\{\underline{p}\}$ give the irreducible grid points of the regular grid. It is
important to perform these operations only by integers for the computational
efficiency.

\subsection{$\mathbf{q}$ point triplets}

In the phono3py code, triplets of $\mathbf{q}$ points,
$(\mathbf{q},\mathbf{q}',\mathbf{q}'')$, are considered for the computation of
the phonon-phonon interaction. This interaction is therefore not a function of a
single $\mathbf{q}$ variable.  The number of grid points $\det(\boldsymbol{D})$
becomes large when dense sampling is used, e.g., $\det(\boldsymbol{D}) =
  300^3=27 \times 10^6$ in the study of Ref.~\onlinecite{IXS-KCl-NaCl}. Then, the
number of triplets of $\mathbf{q}$ points becomes $300^9 \sim 10^{22}$, which is
a really huge number for numerical computations and also storing in memory of
computers.

Due to lattice translational symmetry, elements of the three phonon interaction
strength can be non-zero only if $\mathbf{q} + \mathbf{q}' + \mathbf{q}'' =
  \mathbf{G}$.\cite{Laurent-phph-2011,phono3py} This is used to reduce the
computation of the interactions for all triplets
$(\mathbf{q},\mathbf{q}',\mathbf{q}'') $ to all pairs of points
$(\mathbf{q},\mathbf{q}')$. To satisfy this condition, the $\Gamma$ centred
regular grid is used in the phonon-phonon interaction calculation of the
phono3py code. Moreover, the crystallographic point group operations, time
reversal symmetry, and permutation symmetry for each pair of $\mathbf{q}$ and
$\mathbf{q}'$ are used to reduce further the number of interactions to be
computed.  However, even doing so, the number of the combination of two
$\mathbf{q}$ points is still large. Therefore, in this section,  we describe the
strategy used to make those computations practical.

As mentioned in Sec.\ref{sec:irredubile-grid-points}, many phonon related
properties are written as sum of functions of  a single variable $\mathbf{q}$,
such as $\sum_\mathbf{q} f(\mathbf{q})$. For example, under the relaxation time
approximation, the lattice thermal conductivity (LTC) computed in the phono3py
code can be written as
\begin{align}
  \label{eq:SMRT-kappa}
  \kappa = \frac{1}{NV_\text{c}} \sum_{\mathbf{q}\nu}  \tau_{\mathbf{q}\nu} C_{\mathbf{q}\nu}
  \mathbf{v}_{\mathbf{q}\nu} \otimes \mathbf{v}_{\mathbf{q}\nu},
\end{align}
where $C_{\mathbf{q}\nu}$ and $\mathbf{v}_{\mathbf{q}\nu}$ denote the phonon
mode heat capacity and group velocity, respectively,
\begin{align}
  \label{eq:mode-heat-capacity}
   & C_{\mathbf{q}\nu} =
  k_B \left( \frac{\hbar \omega_{\mathbf{q}\nu}}{k_B T} \right)^2
  \frac{\exp(\hbar \omega_{\mathbf{q}\nu} / k_B T) }
  {[\exp(\hbar \omega_{\mathbf{q}\nu} / k_B T) -1]^2}, \\
   & \mathbf{v}_{\mathbf{q}\nu} =
  \frac{\partial \omega_{\mathbf{q}\nu}}{\partial \mathbf{q}}.
\end{align}
In Eq.~(\ref{eq:mode-heat-capacity}), $\hbar$ and $k_B$ denote the reduced
Planck constant and the Boltzmann constant, respectively, and $T$ is the
temperature. $\tau_{\mathbf{q}\nu}$ in Eq.~(\ref{eq:SMRT-kappa}) is the phonon
lifetime, and the reciprocal of $\tau_{\mathbf{q}\nu}$ is calculated from the
phonon-phonon interaction strength in the phono3py code. The explicit equation
used for $\tau_{\mathbf{q}\nu}=1/2
  \Gamma_{\mathbf{q}\nu}(\omega_{\mathbf{q}\nu})$ is given at Eq.
\ref{eq:imag-selfenergy}.

The  LTC of Eq.~(\ref{eq:SMRT-kappa}) is conveniently computed iterating over
irreducible $\mathbf{q}$ points. Indeed, the mode heat
capacity and the lifetime have the same symmetry as the phonon band structure,
Eq.~(\ref{eq:eigenvector-rotation}),
\begin{align}
   & C_{\mathcal{S} \mathbf{q\nu}} =C_{\mathbf{q\nu}},        \\
   & \tau_{\mathcal{S} \mathbf{q\nu}} = \tau_{\mathbf{q\nu}},
\end{align}
and the derivative of Eq.~(\ref{eq:eigenvector-rotation}) gives
\begin{align}
   & \mathbf{v}_{\mathcal{S} \mathbf{q}\nu} =
  \mathcal{S} \mathbf{v}_{ \mathbf{q}\nu}.
\end{align}
Therefore, the summation in Eq.~(\ref{eq:SMRT-kappa}) can be reduced to a
summation over the irreducible grid points. Denoting $\underline{\mathbf{q}}$
the irreducible points,
% \textcolor{red}{I like this notation, can we used it in
% the section "Irreducible grid points" for p ? instead of $p^{ir}$ ?}
we obtain
\begin{align}
  \label{eq:kappa-sum-ir-grid}
  \kappa = \frac{1}{NV_\text{c}}
  \sum_{\underline{\mathbf{q}}\nu}
  \tau_{\underline{\mathbf{q}}\nu}C_{\underline{\mathbf{q}}\nu} \Big(
  \sum_{\mathcal{S}} \mathcal{S} \mathbf{v}_{\underline{\mathbf{q}}\nu}  \otimes
  \mathcal{S} \mathbf{v}_{\underline{\mathbf{q}}\nu}  \Big)
  \frac{m({\underline{\mathbf{q}}})}{|\{\mathcal{S}\}|}
\end{align}
where the last factor is the number of branches in the star of
$\underline{\mathbf{q}}$, $m({\underline{\mathbf{q}}})$, divided by the
cardinality of the crystallographic point group, $|\{\mathcal{S}\}|$.

When each calculation of $\tau_{\underline{\mathbf{q}}\nu}$ is computationally
demanding, the computation of $\tau_{\underline{\mathbf{q}}\nu}$ at different
$\underline{\mathbf{q}}$ points may be distributed over multiple or many
computer nodes. Therefore, it is convenient to have a set of $\mathbf{q}$ point
triplets at fixed $\underline{\mathbf{q}}$ in $\{(\underline{\mathbf{q}},
  \mathbf{q}', \mathbf{q}'')\}$. As shown in Eq.~(\ref{eq:kappa-sum-ir-grid}), in
this strategy, $\underline{\mathbf{q}}$ is chosen from the irreducible BZ. Now,
because $\underline{\mathbf{q}}$ must be kept fixed, it is not the full set of
operation of the crystallographic point group which should be used to reduce the
summation over $\mathbf{q}'$, but a subgroup which lets $\underline{\mathbf{q}}$
invariant. This subgroup is known as the point group of
$\underline{\mathbf{q}}$,
\begin{align}
  \mathbb{P}_{\underline{\mathbf{q}}} = \{
  \mathcal{S}|
  \mathcal{S}\underline{\mathbf{q}}=\underline{\mathbf{q}} \pmod{\boldsymbol{G}},
  \; \mathcal{S} \in \mathbb{P} \}.
\end{align}
Consequently, $\mathbf{q}'$ is chosen in the irreducible part of the BZ defined
by $\mathbb{P}_{\underline{\mathbf{q}}} $. Finally, $\mathbf{q}''$ is computed
as $\mathbf{q}'' = \mathbf{G} - \underline{\mathbf{q}} - \mathbf{q}'$  where
$\mathbf{G}$ is a reciprocal lattice vector chosen to shift $\mathbf{q}''$
within the BZ of the same origin. For phonons, one of $(\underline{\mathbf{q}},
  \mathbf{q}', \mathbf{q}'')$ and $(\underline{\mathbf{q}}, \mathbf{q}'',
  \mathbf{q}')$ is chosen due to the permutation symmetry.

In the phono3py code, $\mathbf{q}$, $\mathbf{q}'$, and $\mathbf{q}''$ are taken
from the BZ of the same origin. When some of them are on
the BZ surface, they are chosen among their translationally equivalent points on
the BZ surface so that the triplet minimizes $|\mathbf{G}|$ of
$\underline{\mathbf{q}} + \mathbf{q}' + \mathbf{q}'' = \mathbf{G}$.

\begin{figure}[ht]
  \begin{center}
    \includegraphics[width=1.0\linewidth]{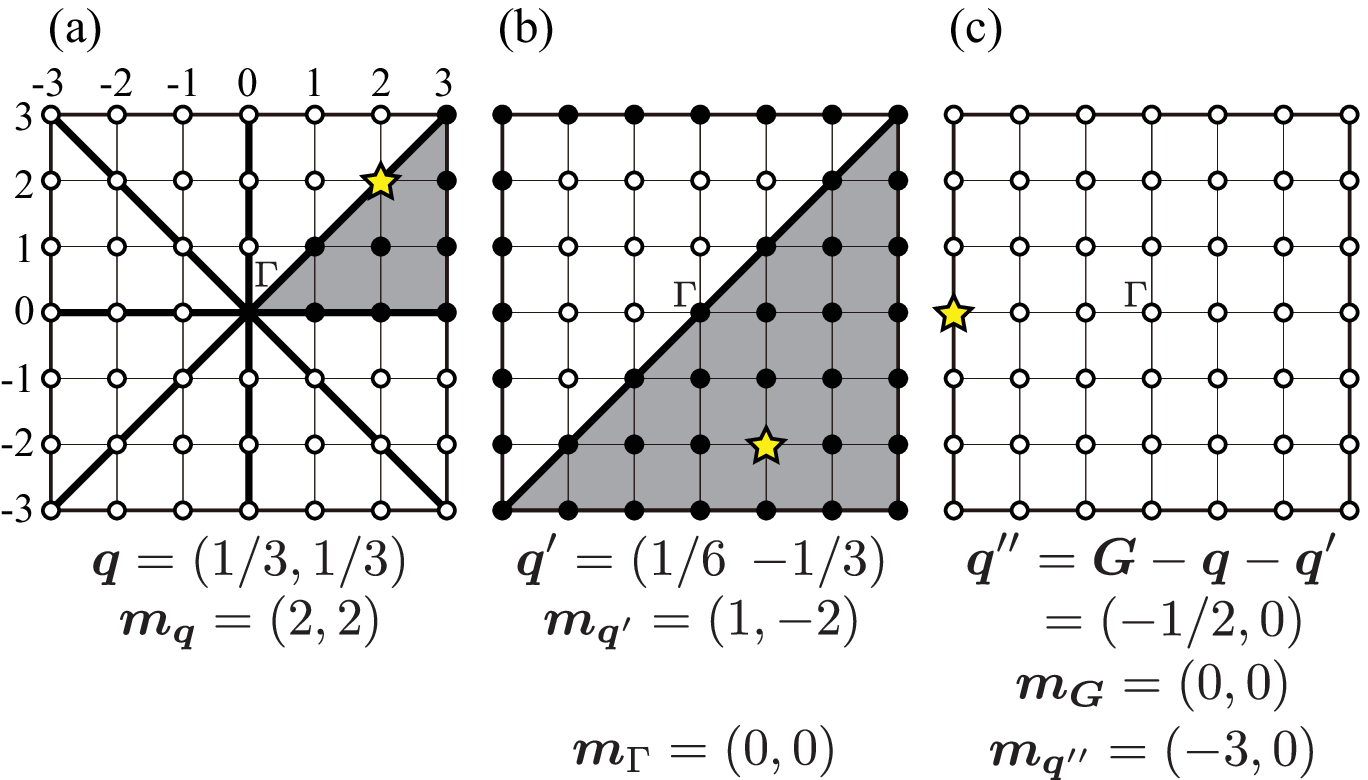}
    \caption{A strategy to choose a $\mathbf{q}$ point triplet. The same 2D
      Brillouin zone (BZ) as Fig.~\ref{fig:grid-symmetry} is used. The star
      symbols depict three $\mathbf{q}$ points in the BZ under the constraint of
      $\boldsymbol{q} + \boldsymbol{q}' + \boldsymbol{q}'' = \boldsymbol{G}$.
      (a) $\boldsymbol{q}$ is chosen from the irreducible grid points of
      $C_\text{4v}$ symmetry. $\boldsymbol{q}$ works as the stabilizer of the
      subgroup of $C_\text{4v}$ symmetry. (b) $\boldsymbol{q}'$ is chosen in the
      irreducible grid points of the subgroup. (c) $\boldsymbol{q}'' =
        \boldsymbol{G} - \boldsymbol{q} - \boldsymbol{q}'$ where $\boldsymbol{G}$
      of the shortest $|(\mathbf{a}^*, \mathbf{b}^*)\boldsymbol{G}|$, i.e.,
      $\boldsymbol{G}=(0, 0)$ in this example, is chosen.
      \label{fig:triplets-symmetry} }
  \end{center}
\end{figure}

This triplet search is implemented using the irreducible grid points and the BZ
grid points described in Secs.~\ref{sec:irredubile-grid-points} and
\ref{sec:grid-points-1st-BZ}, respectively. The $\mathbf{q}$ points
$\underline{\boldsymbol{q}}$, $\boldsymbol{q}'$, and $\boldsymbol{q}''$ are
represented by the BZ grid points
$\boldsymbol{m}^\text{BZ}_{\underline{\boldsymbol{q}}}$,
$\boldsymbol{m}^\text{BZ}_{\boldsymbol{q}'}$, and
$\boldsymbol{m}^\text{BZ}_{\boldsymbol{q}''}$, respectively. As shown in
Fig.~\ref{fig:triplets-symmetry} (a),
$\boldsymbol{m}^\text{BZ}_{\underline{\boldsymbol{q}}}$ is chosen in the
irreducible grid points. The point
$\boldsymbol{m}^\text{BZ}_{\underline{\boldsymbol{q}}}$ breaks the
crystallographic point-group $\mathbb{P}$. Using $
  \tilde{\boldsymbol{S}}^{-\intercal}$ given by Eq.~(\ref{S_tilde}), the point
group of $\underline{\boldsymbol{q}}$ is obtained as
\begin{align}
  \label{eq:rotation-BZ-grid}
  \mathbb{P}_{\underline{\boldsymbol{q}}} = \{
  \mathcal{S}|
  \boldsymbol{m}^\text{BZ}_{\underline{\boldsymbol{q}}} =
  \tilde{\boldsymbol{S}}^{-\intercal}
  \boldsymbol{m}^\text{BZ}_{\underline{\boldsymbol{q}}} \pmod{\boldsymbol{n}},
  \;\; \mathcal{S} \in \mathbb{P} \}.
\end{align}
As shown in Fig.~\ref{fig:triplets-symmetry} (b),
$\boldsymbol{m}^\text{BZ}_{\boldsymbol{q}'}$ is sampled from the irreducible
grid points of $\mathbb{P}_{\underline{\boldsymbol{q}}}$. The third grid point
is given by $\boldsymbol{m}^\text{BZ}_{\boldsymbol{q}''} = \boldsymbol{D}
  \boldsymbol{G} - \boldsymbol{m}^\text{BZ}_{\underline{\boldsymbol{q}}} -
  \boldsymbol{m}^\text{BZ}_{\boldsymbol{q}'}.$ Finally,
$\boldsymbol{m}^\text{BZ}_{\underline{\boldsymbol{q}}}$,
$\boldsymbol{m}^\text{BZ}_{\boldsymbol{q}'}$, or
$\boldsymbol{m}^\text{BZ}_{\boldsymbol{q}''}$ may be shifted to minimize
$|\mathbf{G}|$ if some of them are on the BZ surface as explained for
$(\underline{\mathbf{q}}, \mathbf{q}', \mathbf{q}'')$ above.

\section{Tetrahedron method}
\label{sec:tetrahedron-method}
Once implemented, tetrahedron methods are easy to use and robust. Moreover, when
the code is written in a modular way, it can be reusable for different kind BZ
integrations. The phonopy and phono3py codes employ a linear tetrahedron method
where several techniques are picked and mixed from the various reports, notably,
by MacDonald {\it et al.}\cite{MacDonald-tetrahedron-1979} and Bl\"ochl {\it et
al.}\cite{Blochl-tetrahedron-1994}

\begin{figure}[ht]
  \begin{center}
    \includegraphics[width=0.55\linewidth]{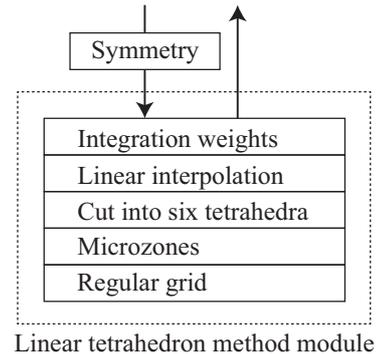}
    \caption{Implementation of the linear tetrahedron method in the phonopy and
      phono3py codes. This routine is provided as a module that
      returns integration weights. In the module, each layer
      depends on the lower layers. Crystal symmetry is treated outside of this
      module.  \label{fig:tetrahedron-structure} }
  \end{center}
\end{figure}

\begin{figure}[ht]
  \begin{center}
    \includegraphics[width=0.9\linewidth]{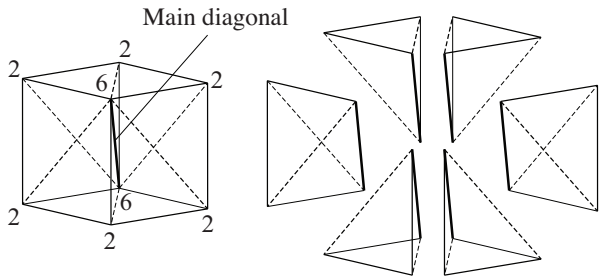}
    \caption{A scheme to divide a parallelepiped microzone into six tetrahedra
      with the same volume.\cite{Blochl-tetrahedron-1994} The eight vertices of
      the parallelepiped are shared by 6, 6, 2, 2, 2, 2, 2, and 2 tetrahedra,
      respectively. \label{fig:tetrahedra-division} }
  \end{center}
\end{figure}

The implementation in the phonopy and phono3py codes is structured as
schematically illustrated in Fig.~\ref{fig:tetrahedron-structure}. The bottom of
the structure is the regular grid. The parallelepiped formed by the basis
vectors of the reciprocal primitive cell is divided into many smaller
parallelepiped microzones on the regular grid. Every microzone has the same
shape, defined by the basis vectors $(\mathbf{a}_\text{m}^*,
  \mathbf{b}_\text{m}^*, \mathbf{c}_\text{m}^*)$, as presented in
Sec.~\ref{sec:regular-grids}. In the case of the traditional regular grid, Eq.
\ref{eq:simple-microzone} defines the microzones to be considered, while for the
generalized regular grid Eq. \ref{eq:generaized-microzone} is used. Each of the
microzones is divided into six tetrahedra in the same way as shown in
Fig.~\ref{fig:tetrahedra-division}.

The integral over the BZ to be performed is then regarded as a sum of
contributions from those tetrahedra. The function to be integrated is linearly
interpolated within each tetrahedron (see Sec.~\ref{sec:tetraherdon-functions})
and therefore the integration within each tetrahedron can be done analytically.
To interpolate linearly a tridimensional function one needs four values. Each
tetrahedron has four vertices. Therefore the values of the function to be
integrated on the vertices of the tetrahedra are used to build the
interpolations.

Different recipes can be used to implement this program. We follow
Ref.~\onlinecite{MacDonald-tetrahedron-1979} as well as
Ref.~\onlinecite{Blochl-tetrahedron-1994} to obtain integration weights for grid
points rather than for the tetrahedra themselves. Indeed, as it will be shown,
it is possible to rearrange the contributions from tetrahedra to contributions
from the grid points. This is useful to make the tetrahedron method easy use,
and to make it similar to smearing methods. In addition, this way, the symmetry
of the regular grid described in Sec.~\ref{sec:symmetry-of-regular-grid} is
applied straightforwardly to the integration weights.

\subsection{Division into six tetrahedra}

The scheme to divide the microzone into six tetrahedra follows
Ref.~\onlinecite{Blochl-tetrahedron-1994} reported by Bl\"ochl {\it et al.}
We choose a shortest main diagonal of the parallelepiped $ (\mathbf{a}_\text{m}^*,
  \mathbf{b}_\text{m}^*, \mathbf{c}_\text{m}^*)$. Then the six
tetrahedra are selected sharing this main diagonal as shown in
Fig.~\ref{fig:tetrahedra-division}.

\subsection{Functions}
\label{sec:tetraherdon-functions}
In this section we introduce the functions  $g(\omega)$, $n(\omega)$,
$I(\omega)$, and $J(\omega)$ to be computed using the linear tetrahedron method.
The notation roughly follows Ref.~\onlinecite{MacDonald-tetrahedron-1979}.
Moreover, the band index $\nu$ is not written explicitly, since each band can be
treated independently. It is therefore understood that a sum over the bands
should be performed at the end of the calculation.

The density of states $g(\omega)$ is written as
\begin{align}
  g(\omega) & \equiv \int_{BZ} d^3q \delta(\omega-\omega_{\mathbf{q}})               \\
            & \approx V_\mathrm{t} \sum^{6N}_{i=1} g(\omega, \omega^i_1, \omega^i_2,
  \omega^i_3, \omega^i_4) \nonumber                                                  \\
            & \equiv V_\mathrm{t} \sum^{6N}_{i=1} g^i.
\end{align}
The definition is written at the first line, the approximation obtained from the
linear tetrahedron method at the second line and an obvious definition of $
  g^i$, the contribution of tetrahedron $i$, at the third line. Here $i$ is an
index running throughout the $6N$ tetrahedra, and $V_\mathrm{t}$ is the volume
of a tetrahedron. The values of frequency at the vertices are assumed to be
arranged in ascending order, $\omega^i_1 < \omega^i_2 < \omega^i_3 <\omega^i_4$.

The integrated density of states, or number of states function, $n(\omega)$,
is written following the same conventions,
\begin{align}
  n(\omega) & \equiv \int^\omega_{-\infty} d\omega'g(\omega')                         \\
            & \approx V_\mathrm{t}  \sum^{6N}_{i=1} n(\omega, \omega^i_1, \omega^i_2,
  \omega^i_3, \omega^i_4)  \nonumber                                                  \\
            & \equiv V_\mathrm{t} \sum^{6N}_{i=1} n^i.
\end{align}

Weighted density of states frequently appears in phonon calculations. They are
defined by
\begin{align}
  \label{eq:bz-integration}
  I(\omega) & = \int_\text{BZ} d^3q F(\mathbf{q})\delta(\omega -\omega_{\mathbf{q}}) \nonumber \\
            & \approx  V_\mathrm{t} \sum^{6N}_{i=1} g^i \sum^4_{k=1} I_k(\omega, \omega^i_1,
  \omega^i_2, \omega^i_3, \omega^i_4) F^i_k  \nonumber                                         \\
            & \equiv V_\mathrm{t} \sum^{6N}_{i=1}  g^i \sum^4_{k=1} I^i_k F^i_k,
\end{align}
where $F(\mathbf{q})$ is a function of  $\mathbf{q}$ and we used the notation
$F^i_k=F(\mathbf{q}^i_k)$ at the vertex $k$ of the tetrahedron labeled by $i$.
$I^i_k=I_k(\omega, \omega^i_1, \omega^i_2, \omega^i_3, \omega^i_4)$ are given in
appendix \ref{tetra_fc}. We notice in this equation an additional summation over
the vertices of each tetrahedron. This comes from the function $F$, which is
linearly interpolated within each tetrahedron. When $F=1$, this equation reduces
to the density of states.

Finally,  the integral of $I(\omega)$ over frequency, $J(\omega)$, is written
as
\begin{align}
  \label{eq:tetrahedron-J-omega}
  J(\omega) & = \int^\omega_{-\infty} d\omega' I(\omega')
  = \int_\mathrm{BZ}
  d^3q F(\mathbf{q})\theta(\omega -\omega_{\mathbf{q}})
  \nonumber                                                              \\
            & \approx V_\mathrm{t}
  \sum^{6N}_{i=1} n^i \sum^4_{k=1} J_k(\omega, \omega^i_1, \omega^i_2, \omega^i_3,
  \omega^i_4) F^i_k
  \nonumber                                                              \\
            & \equiv V_\mathrm{t} \sum^{6N}_{i=1} n^i \sum^4_{k=1} J^i_k
  F^i_k,
\end{align}
where $J^i_k=J_k(\omega, \omega^i_1, \omega^i_2, \omega^i_3,
  \omega^i_4)$ are given in appendix \ref{tetra_fc}.

\begin{figure}[ht]
  \begin{center}
    \includegraphics[width=1.0\linewidth]{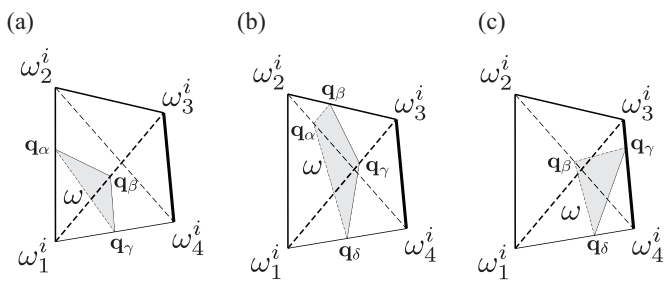}
    \caption{Different linear cuts by $\omega$ plane of a tetrahedron. (a)
      $\omega_1^i < \omega < \omega_2^i < \omega_3^i < \omega_4^i$, (b)
      $\omega_1^i < \omega_2^i < \omega < \omega_3^i < \omega_4^i$, and (c)
      $\omega_1^i < \omega_2^i < \omega_3^i < \omega
        <\omega_4^i$.\cite{MacDonald-tetrahedron-1979}
      \label{fig:cut-tetrahedron} }
  \end{center}
\end{figure}

In appendix \ref{tetra_fc}, the formulae for the coefficients appearing for the
quantities, $g^i$, $n^i$,  $I^i_k$, $J^i_k$, are stated from
Ref.~\onlinecite{MacDonald-tetrahedron-1979}. To derive those formulae, the
calculation of $J(\omega)$ is considered first, and $I(\omega)$ is obtained
through differentiation. $n(\omega)$ and $g(\omega)$ are just special cases for
$F=1$. The $\theta(\omega -\omega_{\mathbf{q}})$ function in $J(\omega)$ shows
that computing the contribution from one tetrahedron is related to the
estimation of the volume $\omega_{\mathbf{q}} < \omega$, which can be described
from an intersection of a tetrahedron with a plane $\omega_{\mathbf{q}} =
  \omega$, since $\omega_{\mathbf{q}}$ is now assumed to be a linear function of
$\mathbf{q}$. As shown in Fig. \ref{fig:cut-tetrahedron} the volume will assume
different shapes depending on the value of $\omega$. Different formulae are
therefore obtained for the different cases to be considered.

\subsection{Integration weights}
\label{sec:integration-weights}

In  Eq.~(\ref{eq:bz-integration}), $I(\omega)$ is expressed as a summation over
$6N$ tetrahedra and four vertices. Following Bl\"ochl {\it et al.} in
Ref.~\onlinecite{Blochl-tetrahedron-1994}, this can be rearranged as a sum over
grid points. This scheme hides cumbersome handling of
tetrahedra from outside the linear tetrahedron method module as shown in
Fig.~\ref{fig:tetrahedron-structure}. Once integration weights, that are
described in this section, are computed, the data structure is shared with
smearing methods and the symmetry handling is made easy.

$I(\omega)$  can be rewritten as
\begin{align}
  \label{eq:24-tetrahedra}
  I(\omega) \approx V_\mathrm{t} \sum^{N-1}_{p=0}\sum^{24}_{i_p=1} g^{i_p}
  \sum^4_{k=1} I^{i_p}_k F^{i_p}_k \delta_{(i_p,k),p},
\end{align}
where $p$ denotes the grid points (see Sec.~\ref{sec:grid-indexing}), and $i_p$
is the index of a tetrahedron in the 24 tetrahedra that are sharing the grid
point $p$ (see Fig.~\ref{fig:tetrahedra-division} for
$24=6+6+2+2+2+2+2+2$). The composite index $(i_p, k)$
indicates a grid point, and $\delta_{(i_p,k),p}$ selects the vertex which is
identical to the grid point $p$ in the four vertices of the tetrahedron $i_p$.

Since $F^{i_p}_k \delta_{(i_p,k),p}$ is only non zero when $(i_p,k)$ is the
point $p$, we write  $F^{i_p}_k \delta_{(i_p,k),p}= F_p \delta_{(i_p,k),p}$. The
summation~(\ref{eq:24-tetrahedra}) becomes
\begin{align}
  \label{eq:tetrahedron-integration}
  I(\omega) \approx
  V_\mathrm{t} \sum^{N-1}_{p=0} F_p w_p,
\end{align}
where
\begin{align}
  w_p = \sum^{24}_{i_p=1} g^{i_p} \sum^4_{k=1} I^{i_p}_k
  \delta_{(i_p,k),p}. \label{eq:integration-weights}
\end{align}
$ I(\omega) $ has therefore been expressed as a summation over grid points. This
integral is given by the weighted sum of the values of the function $F$ at
points $p$, $F_p$, with weights $w_p$ given by
Eq.~(\ref{eq:integration-weights}).

\begin{figure}[ht]
  \begin{center}
    \includegraphics[width=0.9\linewidth]{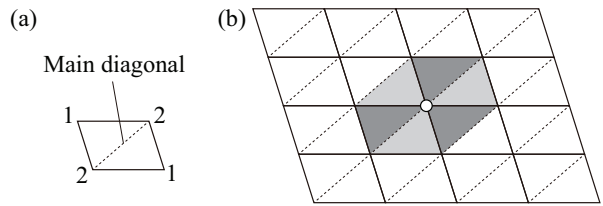}
    \caption{2D schematic illustration of rearrangement of
      summation~(\ref{eq:24-tetrahedra}). (a) Each parallelogram is cut into two
      triangles. Four vertices of the parallelogram are shared by 2, 2, 1, and 1
      triangles, respectively. (b) Six triangles (shaded) that share a grid
      point (circle symbol) contribute to the integration weight of the grid
      point. \label{fig:rearrange-tetrahedra} }
  \end{center}
\end{figure}

A 2D illustration of the summation~(\ref{eq:tetrahedron-integration}) is
presented in Fig.~\ref{fig:rearrange-tetrahedra}.
Fig.~\ref{fig:rearrange-tetrahedra} (a) shows that each of parallelograms is cut
in two triangles, and the four vertices of each parallelogram are shared by 2,
2, 1, and 1 of those triangles. Six triangles share the grid point marked by the
circle symbol as shown in Fig.~\ref{fig:rearrange-tetrahedra} (b). This shows
that only 6 terms are summed up in the summation~(\ref{eq:integration-weights})
because of $\delta_{(i_p,k),p}$. For the tetrahedra in 3D, 24 terms are summed
up at each grid point in the summation~(\ref{eq:integration-weights}).

\begin{figure}[ht]
  \begin{center}
    \includegraphics[width=0.6\linewidth]{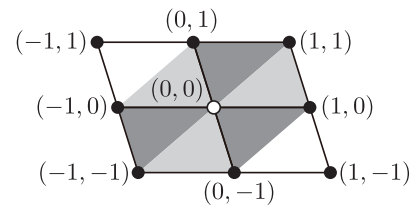}
    \caption{2D schematic illustration of grid point shifts $\Delta
        \boldsymbol{m}$ from the focused grid point (center) to the neighbors.
      \label{fig:relative-grid-points} }
  \end{center}
\end{figure}

To compute Eq.~(\ref{eq:integration-weights}), it is convenient to represent the
positions of grid points using shifts from the grid point $p$, $\boldsymbol{m}_p
+ \Delta \boldsymbol{m}$. The 2D example is depicted in
Fig.~\ref{fig:relative-grid-points}. The main diagonal $(0, 0) \text{--} (1, 1)$
is chosen. For this main diagonal, the six triangles that share the central
point are uniquely determined. They are represented by the following $\Delta
\boldsymbol{m}$ of the apices: $(0, 0) \text{--} (0, 1) \text{--} (1, 1), (0, 0)
\text{--} (1, 1) \text{--} (1, 0), \ldots$

For the linear tetrahedron method, there are four choices for the main diagonal,
and for each choice of the main diagonal, the $\Delta \boldsymbol{m}$ of the
vertices of the 24 tetrahedra shared by a grid point are predetermined.
Therefore, the datasets of the 24 tetrahedra of the four main diagonals are
predetermined and hard-corded in the phonopy and phono3py codes as $4\times
  24\times 4 \times 3$ integer array whose elements are in $\{-1, 0, 1\}$ for the
regular grid.

For the generalized regular grid, the shifts -1 or 1 corresponds to coordinates
of the neighboring points in the $(\mathbf{a}_\text{m}^*, \mathbf{b}_\text{m}^*,
  \mathbf{c}_\text{m}^*)$ basis. The coordinates of those shifts are however
easily obtained in the $(\tilde{\mathbf{a}}_\text{m}^*,
  \tilde{\mathbf{b}}_\text{m}^*,  \tilde{\mathbf{c}}_\text{m}^*)$ basis used in
the calculation,  $\Delta \boldsymbol{m}^\text{G} = \boldsymbol{P} \Delta
  \boldsymbol{m}$, since
\begin{align}
  \label{eq:relative-shifts-Cart}
  (\mathbf{a}_\text{m}^*, \mathbf{b}_\text{m}^*,
  \mathbf{c}_\text{m}^*) \Delta \boldsymbol{m}
   & =
  (\mathbf{a}_\text{m}^*, \mathbf{b}_\text{m}^*,
  \mathbf{c}_\text{m}^*) \boldsymbol{P}^{-1}
  \boldsymbol{P} \Delta \boldsymbol{m}
  \nonumber \\
   & =
  (\tilde{\mathbf{a}}_\text{m}^*, \tilde{\mathbf{b}}_\text{m}^*,
  \tilde{\mathbf{c}}_\text{m}^*)\Delta \boldsymbol{m}^\text{G},
\end{align}
where Eq.~\ref{eq:rec-basis-vectors-P} is used in the last equation.

\subsection{Use of symmetry of grid points}

The symmetry of grid points can be applied to the linear tetrahedron method.
The symmetry is used to prepare input dataset of a grid point
required by the linear tetrahedron method module as illustrated in
Fig.~\ref{fig:tetrahedron-structure}. Once the input dataset is made, symmetry
information is unnecessary for the linear tetrahedron method module.

In this section, we denote $\omega_p \equiv \omega_k^{i}$ and the mapping of the
grid point $p \xrightarrow[]{\{\mathcal{S}\}} \underline{p}$ defined in
Sec.~\ref{sec:irredubile-grid-points} is written as $\underline{p}=f(p)$. Under
the condition $\omega_p = \omega_{f(p)}$, which is satisfied by the phonon
frequencies, the integration weights fulfill the relation
\begin{align}
  \label{eq:ir-integration-weights}
  w_p & \approx w_{f(p)}.
\end{align}
This is an approximation since the set of
24 tetrahedra may not follow the crystallographic point group, i.e., the
rotated tetrahedron may not be mapped to any of the 24 tetrahedra. However,
Eq.~(\ref{eq:ir-integration-weights}) is a very good approximation.

When $F_p$ has the symmetry $F_p = F_{f(p)}$,
Eq.~(\ref{eq:tetrahedron-integration}) can be written as
\begin{align}
  \label{eq:tetrahedron-ir-integration}
  I(\omega) \approx V_\mathrm{t} \sum^{N-1}_{p=0} F_p w_p \approx
  V_\mathrm{t} \sum^{N-1}_{p=0} F_{f(p)} w_{f(p)}.
\end{align}
The right hand side of Approx.~(\ref{eq:tetrahedron-ir-integration}) can
therefore be computed from the values at irreducible grid points only. Defining
the multiplicity of the irreducible points as
$m(\underline{p})=|\{p|\underline{p}=f(p), \; \forall p\}|$, the right hand side
of Approx.~(\ref{eq:tetrahedron-ir-integration}) is written as
\begin{align}
  I(\omega) \approx
  V_\mathrm{t} \sum_{\{\underline{p}\}} F_{\underline{p}} w_{\underline{p}}
  m(\underline{p}).
\end{align}
Here $ m(\underline{p})$ is equivalent to $m(\underline{\mathbf{q}})$ in
Eq.~(\ref{eq:kappa-sum-ir-grid}), i.e. the number of branches in the star of
$\underline{\mathbf{q}}$.

\subsection{Triplets integration weights}

As shown in Ref.~\onlinecite{phono3py}, the phonon linewidth computed by the
phono3py code, due to phonon-phonon interactions, is given by
\begin{widetext}
  \begin{align}
    \label{eq:imag-selfenergy}
    \Gamma_{\mathbf{q}\nu}(\omega) & = \frac{18\pi}{\hbar^2}
    \sum_{\mathbf{q}' \mathbf{q}''}  \sum_{\nu' \nu''}
    \bigl|\Phi_{-\mathbf{q}\nu,\mathbf{q}'\nu',\mathbf{q}''\nu''}\bigl|^2
    \left\{(n_{\mathbf{q}'\nu'}+ n_{\mathbf{q}''\nu''}+1)
    \delta(\omega-\omega_{\mathbf{q}'\nu'}-\omega_{\mathbf{q}''\nu''}) \right.     \\
                                   & + (n_{\mathbf{q}'\nu'}-n_{\mathbf{q}''\nu''})
    \left[\delta(\omega+\omega_{\mathbf{q}'\nu'}-\omega_{\mathbf{q}''\nu''})
      - \left. \delta(\omega-\omega_{\mathbf{q}'\nu'}+\omega_{\mathbf{q}''\nu''})
      \right]\right\}.
  \end{align}
\end{widetext}
If we consider $\mathbf{q}$ to be fixed, the summations to be performed over the
BZ have the form
\begin{align}
  K_{\mathbf{q}}(\omega) & = \frac{1}{N}
  \sum_{\mathbf{q}' \mathbf{q}''} F_{\mathbf{q}, \mathbf{q}', \mathbf{q}''}
  \Delta(\mathbf{q} + \mathbf{q}' + \mathbf{q}'')
  \delta(\omega-  \omega_{\mathbf{q}', \mathbf{q}''} )
\end{align}
where $ F_{\mathbf{q} ,\mathbf{q}' ,\mathbf{q}''}$ is an arbitrary function of $
\mathbf{q}$, $\mathbf{q}'$ and $\mathbf{q}''$. $\omega_{\mathbf{q}',
\mathbf{q}''} $ is a function of $\mathbf{q}'$ and $\mathbf{q}''$. For example,
in Eq. \ref{eq:imag-selfenergy}, to compute the contribution from the first
delta function we need to consider, $\omega_{\mathbf{q}', \mathbf{q}''} =
\omega_{\mathbf{q}'} +  \omega_{\mathbf{q}''} $.

In the above equation, $\Delta(\mathbf{q} + \mathbf{q}' + \mathbf{q}'')$ means 1
when $\mathbf{q} + \mathbf{q}' + \mathbf{q}'' = \mathbf{G}$ otherwise 0.  In
Eq.~\ref{eq:imag-selfenergy} this factor is included in $
\Phi_{\mathbf{q}\nu,\mathbf{q}'\nu',\mathbf{q}''\nu''}$.\cite{phono3py}  For
given $\mathbf{q}$, $\mathbf{q}'$, and $\mathbf{q}''$ in the BZ, $\mathbf{q}''$
is chosen so that $|\mathbf{G}|$ is smallest. At fixed $\mathbf{q}$, for given
$\mathbf{q}'$, only a single $\mathbf{q}''$ contributes. This allows to
eliminate the summation over $\mathbf{q}''$, therefore we obtain

\begin{align}
  \label{eq:integral-of-three-phonon-function}
  K_{\mathbf{q}}(\omega) & = \frac{1}{N} \sum_{\mathbf{q}' }
  F_{\mathbf{q}, \mathbf{q}', \mathbf{G} - \mathbf{q} -  \mathbf{q}'}
  \delta(\omega-  \omega_{\mathbf{q}', \mathbf{G} - \mathbf{q} -  \mathbf{q}' } ) \\
                         & = \frac{V_{\text{c}} }{(2\pi)^3} \int_\text{BZ} d^3 q'
  F_{\mathbf{q}, \mathbf{q}', \mathbf{G} - \mathbf{q} -  \mathbf{q}'}
  \delta(\omega-  \omega_{\mathbf{q}', \mathbf{G} - \mathbf{q} -  \mathbf{q}' } ).
\end{align}
This
quantity has therefore the form of the function $I(\omega)$ we considered
previously. To apply the linear tetrahedron method to $K_{\mathbf{q}}(\omega)$
we need to know the value of the functions $ F_{\mathbf{q}, \mathbf{q}',
      \mathbf{G} - \mathbf{q} - \mathbf{q}'} $ and $\omega_{\mathbf{q}', \mathbf{G}
    - \mathbf{q} - \mathbf{q}' }$ at the vertices $\mathbf{q}' + \Delta
  \mathbf{q}'$ around the point $\mathbf{q}'$. They are given by
\begin{align}
   & F_{\mathbf{q}, \mathbf{q}'+\Delta \mathbf{q}', \mathbf{G} - \mathbf{q} -
      \mathbf{q}'-\Delta \mathbf{q}'}=F_{\mathbf{q}, \mathbf{q}' +
  \Delta \mathbf{q}', \mathbf{q}'' -\Delta \mathbf{q}'},                      \\
   & \omega_{\mathbf{q}'+\Delta \mathbf{q}', \mathbf{G} - \mathbf{q} -
    \mathbf{q}'-\Delta \mathbf{q}' }=\omega_{\mathbf{q}'+\Delta \mathbf{q}',
    \mathbf{q}'' -\Delta \mathbf{q}' }.
\end{align}
It means that  $\Delta \mathbf{q}'' = -\Delta \mathbf{q}'$ for the corresponding
neighboring point of $\mathbf{q}''$. Using the values
$\omega_{\mathbf{q}'+\Delta \mathbf{q}', \mathbf{q}'' -\Delta \mathbf{q}'}$ as
the input dataset of the linear tetrahedron method, the triplets integration
weights are computed in the same way as written in
Sec.~\ref{sec:integration-weights}.

\section{Random displacement generation}
\label{seq:random-displacement}
Once phonon frequency $\omega_{\mathbf{q}\nu}$ and eigenvector
$W_{\kappa\alpha}(\mathbf{q}\nu)$ are obtained in a supercell, they can be used
to generate atomic configurations relevant at finite temperatures. This can be
achieved using Eq.~(\ref{eq:displacement-operator}) and writing
\begin{equation}
  Q(\mathbf{q}\nu) = \frac{1}{\sqrt{2}}(Q^{R}(\mathbf{q}\nu)+iQ^{I}(\mathbf{q}\nu)), \label{QRI}
\end{equation}
where both $Q^{R}(\mathbf{q}\nu)$ and $Q^{I}(\mathbf{q}\nu)$ are real variables.
It is shown in Ref.~\onlinecite{TheoryOfLatticeDynamics} that they are real
normal coordinates,  fulfilling the equation of the harmonic oscillator. Their
probability density is therefore a gaussian,
\begin{equation}
  P[Q^{R/I}(\mathbf{q}\nu)] = \frac{1}{\sqrt{2\pi\sigma^2_{\mathbf{q}\nu}}}
  \exp [-\frac{Q^{R/I}(\mathbf{q}\nu)^2}{2\sigma^2_{\mathbf{q}\nu}}],
  \label{eq:distribution_of_Q}
\end{equation}
with
\begin{equation}
  \label{eq:variance_of_Q}
  \sigma_{\mathbf{q}\nu}^2 =  \frac{\hbar}{2\omega_{\mathbf{q}\nu}} \coth \frac{\hbar \omega_{\mathbf{q}\nu}}{2k_\text{B}T}
  = \frac{\hbar}{2\omega_{\mathbf{q}\nu}}(1+2n_{\mathbf{q}\nu}),
\end{equation}
where $n_{\mathbf{q}\nu} =(e^{\frac{\hbar\omega_{\mathbf{q}\nu}}{k_\text{B}T}}-1)^{-1}$ is the
Bose--Einstein distribution function.

Substituting Eq.~(\ref{QRI}) into Eq.~(\ref{eq:displacement-operator}) and
remembering $Q(-\mathbf{q}\nu)=Q^{*}(\mathbf{q}\nu)$, we obtain
\begin{align}
  u_{l{\kappa}\alpha} & = \frac{1}{\sqrt{Nm_{\kappa}}}
  \Bigg\{ \sum_{\mathbf{q}\in A, \nu}
  Q(\mathbf{q}\nu)
  \Re [u_{l\kappa\alpha}(\mathbf{q}\nu)] \notag                                            \\
                      & + \frac{1}{\sqrt{2}} \sum_{\mathbf{q}\in B, \nu}
  \bigg[(Q^{R}(\mathbf{q}\nu)+iQ^{I}(\mathbf{q}\nu))
  u_{l\kappa\alpha}(\mathbf{q}\nu)  \notag                                                 \\
                      & \hspace{15mm} + (Q^{R}(\mathbf{q}\nu)-iQ^{I}(\mathbf{q}\nu))
  u^{*}_{l\kappa\alpha}(\mathbf{q}\nu) \bigg]\Bigg\}                                       \\
                      & = \frac{1}{\sqrt{Nm_{\kappa}}} \Bigg\{ \sum_{\mathbf{q}\in A, \nu}
  Q(\mathbf{q}\nu)
  \Re [u_{l\kappa\alpha}(\mathbf{q}\nu)] \notag                                            \\
                      & + \sqrt{2} \sum_{\mathbf{q}\in B, \nu}  \bigg[Q^{R}(\mathbf{q}\nu)
  \Re[u_{l\kappa\alpha}(\mathbf{q}\nu)] \notag                                             \\
                      & \hspace{20mm} - Q^{I}(\mathbf{q}\nu)
    \Im[u_{l\kappa\alpha}(\mathbf{q}\nu)] \bigg]\Bigg\}.
  \label{eq:u_by_QR_QI}
\end{align}
Here, $A$ is a set of $\mathbf{q}$ points commensurate with the supercell, where
$Q(\mathbf{q}\nu)$ becomes real (e.g. $\Gamma$ point) or $\mathbf{q}=-\mathbf{q}
+ \mathbf{G}$, and $B$ includes other commensurate $\mathbf{q}$ points in the
one side of the BZ and $\mathbf{q} \neq -\mathbf{q} + \mathbf{G}$. Namely, only
one of $\mathbf{q}$ or $-\mathbf{q}$ should be included in the summation over
$\mathbf{q}\in B$.\cite{TheoryOfLatticeDynamics}

Finally, generating random numbers
following Eq.~(\ref{eq:distribution_of_Q}) and using them for
$Q^{R}(\mathbf{q}\nu)$ and $Q^{I}(\mathbf{q}\nu)$ in
Eq.~(\ref{eq:u_by_QR_QI}), random displacements $\{u_{l\kappa\alpha}\}$ in a
supercell can be generated systematically at a given temperature.

The random structures generated in this way can be used to investigate the
impact of phonon excitation on various physical properties, such as electronic
structure and magnetism, at finite temperatures. Indeed, the investigated
quantity is computed for each random structure generated, and the average of
values obtained for those random structures is computed to mimic the effect of
temperature.

We note that this approach is physically
valid only when all phonon modes are dynamically stable
($\omega_{\mathbf{q}\nu}^2 > 0)$ at all commensurate $\mathbf{q}$ points.
%If an unstable mode exists, phonopy replaces its frequency as $\omega_{\mathbf{q}\nu}\longrightarrow |\omega_{\mathbf{q}\nu}|$ and uses it for Eq.~(\ref{eq:variance_of_Q}).
This condition is usually satisfied in many compounds where lattice
anharmonicity is not very strong. However, in some anharmonic systems, including
high-temperature metastable phases, the harmonic approximation may yield
unstable phonon modes.

Even when all phonons are stable, their frequencies may deviate from
experimental values significantly. In such cases, it is recommended to use
anharmonic phonon frequencies and eigenvectors at finite temperatures for
Eqs.~(\ref{eq:variance_of_Q}) and (\ref{eq:u_by_QR_QI}). The anharmonic
frequencies and eigenvectors renormalized at finite temperatures can be
obtained, for example, by performing a self-consistent
phonon calculation using either the real-space stochastic
approaches,\cite{Monacelli-SSCHA-2021, Errea-SSCHA-2013, Errea-SSCHA-2014,
  Bianco-SSCHA-2017, Monacelli-SSCHA-2018, hiPhive} which are often termed the
stochastic self-consistent harmonic approximation (SSCHA), or the momentum-space
implementation which uses anharmonic force constants.\cite{Tadano-2015,
  Wu.Duan.2020} An application of SSCHA performed with random
displacements generated by the phonopy code coupled with the ALM force constants
calculation code~\cite{Tadano-ALM-2018} is found in
Ref.~\onlinecite{IXS-KCl-NaCl}.

\section{Phonon band unfolding}
\label{seq:phonon-band-unfolding}

To draw a phonon band structure for system with defects computed in a supercell
approach, band unfolding technique may be useful. Since even one point defect
breaks the periodicity of a crystal, a certain approximation is necessary to
achieve this representation. In this section, an implementation of the method
proposed by Allen {\it et al.}~\cite{Allen-band-unfolding} is explained.

\begin{figure}[ht]
  \begin{center}
    \includegraphics[width=1.0\linewidth]{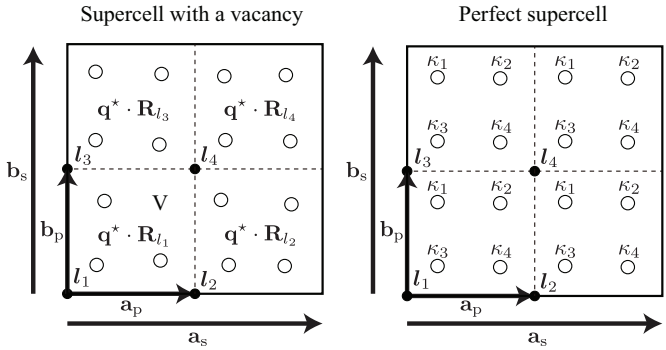}
    \caption{2D schematic illustration of supercell with a vacancy (left) and
      the corresponding perfect supercell (right). The atoms are depicted by
      open circle symbols. The vacancy is marked by "V".  The perfect supercell
      contains $2 \times 2$ primitive cells whose lattice points
      ($\boldsymbol{l}_1, \boldsymbol{l}_2, \boldsymbol{l}_3, \boldsymbol{l}_4$)
      are indicated by filled circle symbols. Positions of all atoms in the
      defective supercell can be deviated slightly from the equilibrium
      positions of the perfect supercell. At unfolding, phase shift of
      $\mathbf{q^\star} \cdot \mathbf{R}_{l_j}$ is multiplied to the phonon
      eigenvector of the defective supercell at each primitive lattice
      translation as given in Eq.(\ref{eq:phonon-unfolding-weight2}).
      \label{fig:unfolding-supercell} }
  \end{center}
\end{figure}

\begin{figure}[ht]
  \begin{center}
    \includegraphics[width=0.45\linewidth]{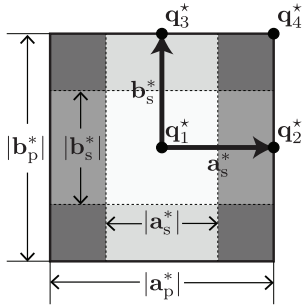}
    \caption{Unfolding of 2D Brillouin zone (BZ) of the supercell
      ($\mathbf{q}^\star_1$) to $2 \times 2$ BZs of the supercell
      ($\mathbf{q}^\star_1, \mathbf{q}^\star_2, \mathbf{q}^\star_3,
        \mathbf{q}^\star_4$). Gray levels of shaded backgrounds indicate BZs of
      the supercell which belong to different $\mathbf{q}^\star$ points.
      \label{fig:unfolding-supercell-BZ} }
  \end{center}
\end{figure}

In Fig.~\ref{fig:unfolding-supercell}, a 2D schematic illustration of a
supercell with a vacancy is presented. It is analyzed by presuming the
corresponding perfect supercell with the same lattice vectors.
Eq.~(\ref{eq:primitive-to-supercell}) gives the relationship between the basis
vectors of the primitive cell and those of the supercell, which indicates that
the perfect supercell contains
$N=\det(\boldsymbol{M}_{\text{p}\rightarrow\text{s}})$ primitive cells. The
reciprocal basis vectors are given by
Eq.~(\ref{eq:reciprocal-primitive-to-supercell}). The BZ volume of the primitive
cell is $N$ times larger than that of the supercell as shown in
Fig.~\ref{fig:unfolding-supercell-BZ}.

For this analysis to be possible, we assume there exits one-to-one
correspondence between the atoms and vacancy sites in the defective supercell
and the atoms in the perfect supercell. The atoms and vacancies in the defective
supercell can therefore be labeled by the composite index $l\kappa$ in the same
way as in the perfect supercell. It means that vacancy sites in the defective
supercell must have respective atoms in the perfect supercell. The atoms in
the defective supercell can nevertheless have slightly different positions than
in the perfect supercell. The 2D schematic illustration we obtain is shown in
Fig.~\ref{fig:unfolding-supercell}.

We write coordinates of lattice points in the supercell as
\begin{align}
  \mathbf{R}_{l}=(\mathbf{a}_\text{p}, \mathbf{b}_\text{p},
  \mathbf{c}_\text{p}){\boldsymbol{l}}
  =(\mathbf{a}_\text{s}, \mathbf{b}_\text{s},
  \mathbf{c}_\text{s})\tilde{\boldsymbol{l}},
\end{align}
where ${\boldsymbol{l}}$ are integers, while $\tilde{\boldsymbol{l}}$ may not
be.

Let $\hat{T}_j$ be the $j$-th lattice translation operator which is defined on a
function as follows
\begin{align}
  \label{eq:translation-of-function}
  \hat{T}_j f(\mathbf{R}) = f(\mathbf{R} - \mathbf{R}_{l_j}).
\end{align}
In this section, $\boldsymbol{l}$ correspond to lattice vectors within the
perfect supercell. There are $N$ such lattice vectors and therefore $N$ lattice
translation operations. $\hat{T}_j$ maps a lattice vector with index $l_i$ in
the perfect supercell to a lattice vector with index $l_k$. The mapping between
indices is determined from
\begin{align}
  \tilde{\boldsymbol{l}}_i \xrightarrow[]{\hat{T}_j} \tilde{\boldsymbol{l}}_k =
  \tilde{\boldsymbol{l}}_i +
  \tilde{\boldsymbol{l}}_j \pmod {\mathbf{1}},
\end{align}
where $\pmod{\mathbf{1}}$ means that the lattice points translated outside the
supercell are brought back inside the supercell by a supercell lattice
translation. The mapping obtained this way is stored in an $N\times N$
permutation matrix.  In Fig.~\ref{fig:unfolding-supercell}, the 2D supercell
model is made of $2\times 2$ primitive cells. For example, the translation by
$\boldsymbol{a}_\text{p}$ brings $l_1$ to $l_2$. One more translation by
$\boldsymbol{a}_\text{p}$ brings $l_2$ to $l_1$ due to the periodicity of the
supercell. In the following, the mapping $l_i \xrightarrow[]{\hat{T}_j} l_k$ is
written compactly as $l_k=T_j(l_i)$.

The first step of this method is to compute the phonon eigenvectors
of the defective supercell. The Bloch wave of an eigenvector given by
Eqs.~(\ref{eq:dynamical-matrix}) and (\ref{eq:phonon-eigenvalue-problem}) is
written as

%\begin{align}
%  \Psi_{\tilde{\mathbf{q}} \tilde{\nu}}( \mathbf{R}_{l \kappa})= \langle \mathbf{R}_{l \kappa} | \Psi_{\tilde{\mathbf{q}} \tilde{\nu}} \rangle = \Psi_{\tilde{\mathbf{q}} \tilde{\nu}}^{l \kappa}
%\end{align}

\begin{align}
  \label{eq:Bloch-wave}
  \Psi_{l{\kappa}}(\tilde{\mathbf{q}} \tilde{\nu}) =
  e^{i\tilde{\mathbf{q}}\cdot\mathbf{R}_{l{\kappa}}}
  \mathbf{W}_{l\kappa}(\tilde{\mathbf{q}} \tilde{\nu}),
\end{align}
where $\mathbf{W}_{l\kappa}(\tilde{\mathbf{q}} \tilde{\nu})$ is the phonon
eigenvector of the defective supercell, with band index $\tilde{\nu}$, and at
the wave vector $\tilde{\mathbf{q}}$ in the BZ of the supercell.
Eigenvector elements corresponding to vacancies are treated as zeros.

We consider that $\hat{T}_j$ is applied to the Bloch wave as
\begin{align}
  \label{eq:translate-Bloch-wave}
  \hat{T}_j\Psi_{l{\kappa}}(\tilde{\mathbf{q}} \tilde{\nu}) & \approx
  \Psi_{T_j^{-1}(l){\kappa}}(\tilde{\mathbf{q}} \tilde{\nu})
  \nonumber
  \\
                                                            &
  =e^{i\tilde{\mathbf{q}} \cdot
  (\mathbf{R}_{l{\kappa}} - \mathbf{R}_{l_j})}
  \mathbf{W}_{T_j^{-1}(l)\kappa}(\tilde{\mathbf{q}} \tilde{\nu}).
\end{align}
The first line of Eq.~(\ref{eq:translate-Bloch-wave}) may not be consistent with
Eq.~(\ref{eq:translation-of-function}) since positions of atoms in the defective
supercell can be displaced slightly from those in the perfect supercell.

The next step is to unfold the $\tilde{\mathbf{q}}$ point from the BZ of the
supercell to the $N$ $\mathbf{q}$ points in the BZ of the primitive cell which
are equivalent up to vectors of the reciprocal supercell lattice. This is
written as
\begin{align}
  \mathbf{q} = \mathbf{q}^\star + \tilde{\mathbf{q}},
\end{align}
where $\mathbf{q}^\star$ is an integer linear combination of
$(\mathbf{a}_\text{s}^{*},  \mathbf{b}_\text{s}^{*}, \mathbf{c}_\text{s}^{*})$.
The schematic illustration for the BZ of the  2D supercell is shown in
Fig.~\ref{fig:unfolding-supercell-BZ}. The $\tilde{\mathbf{q}}$ points in the
BZ of the supercell are shifted by $\mathbf{q}^\star$ inside the BZ
of the primitive cell.

Applying periodic boundary conditions on the supercell, the translation
operations $\{\hat{T}_j\}$ form an Abelian group of order $N$ with irreducible
representations $e^{-i \mathbf{q^\star} \cdot \mathbf{R}_{l_j}}$. The great
orthogonality theorem then gives the relation
\begin{align}
  \frac{1}{N} \sum_{j=1}^{N} e^{i\mathbf{q}^\star \cdot
      \mathbf{R}_{l_j}}=\delta_{\mathbf{q}^\star,0}.
\end{align}
This result is also proved algebraically in appendix \ref{sum_f}, together with
the dual relation
\begin{align}
  \label{eq:completeness-relation}
  \frac{1}{N} \sum_{i=1}^{N} e^{-i\mathbf{q}^\star_i \cdot (
      \mathbf{R}_{l_j} - \mathbf{R}_{l_{j'}})}=\delta_{j, j'}.
\end{align}
where $\mathbf{q}^\star_i $ are the $N$ vectors $\mathbf{q}^\star$ within the BZ
of the primitive cell. We define the operator
\begin{gather}
  \label{eq:unfolding-projector}
  \hat{P}(\tilde{\mathbf{q}}, \mathbf{q^\star}) = \frac{1}{N} \sum_{j=1}^N
  e^{i (\tilde{\mathbf{q}} + \mathbf{q^\star}) \cdot \mathbf{R}_{l_j}} \hat{T}_j .
\end{gather}
It can easily be shown that it is a projection operator, which selects the part
of $\Psi_{l{\kappa}}(\tilde{\mathbf{q}} \tilde{\nu})$ that is also a Bloch
function of the primitive lattice with wave vector $\mathbf{q} = \mathbf{q}^\star
  + \tilde{\mathbf{q}}$. Indeed, in appendix \ref{PC_Bloch} we obtain
\begin{align}
  \hat{T}_i   \hat{P}(\tilde{\mathbf{q}}, \mathbf{q^\star})
  \Psi_{l{\kappa}}(\tilde{\mathbf{q}} \tilde{\nu})
   & =e^{-i (\tilde{\mathbf{q}} + \mathbf{q^\star}) \cdot\mathbf{R}_{l_i}}
  \hat{P}(\tilde{\mathbf{q}}, \mathbf{q^\star})
  \Psi_{l{\kappa}}(\tilde{\mathbf{q}} \tilde{\nu}).
\end{align}
Moreover, because from Eq.~(\ref{eq:completeness-relation}), $\sum_{i=1}^N
  \hat{P}(\tilde{\mathbf{q}}, \mathbf{q}_i^\star) = 1$, the decomposition of
$\Psi_{l{\kappa}}(\tilde{\mathbf{q}} \tilde{\nu})$ into the $N$ functions
$\hat{P}(\tilde{\mathbf{q}}, \mathbf{q}_i^\star)
  \Psi_{l{\kappa}}(\tilde{\mathbf{q}} \tilde{\nu}), i=1,\ldots,N$ is unique,
\begin{gather}
  \label{eq:band-unfolding-decomposition}
  \Psi_{l{\kappa}}(\tilde{\mathbf{q}}
  \tilde{\nu}) = \sum_{i=1}^N
  \left[\hat{P}(\tilde{\mathbf{q}}, \mathbf{q}^\star_i)
    \Psi_{l{\kappa}}(\tilde{\mathbf{q}}
    \tilde{\nu})\right].
\end{gather}

The projection by Eq.~(\ref{eq:unfolding-projector}) and the decomposition
by Eq.~(\ref{eq:band-unfolding-decomposition}) indicate that the weight of state
having the Bloch symmetry $ \mathbf{q} = \mathbf{q}^\star + \tilde{\mathbf{q}}$
of the primitive lattice in the supercell Bloch state
$\Psi_{l_i{\kappa}}(\tilde{\mathbf{q}} \tilde{\nu})$ is therefore defined as
\begin{align}
  \label{eq:phonon-unfolding-weight1}
  w(\tilde{\mathbf{q}}\tilde{\nu}, \mathbf{q}^\star) & = \sum_{i\kappa}
  |\hat{P}(\tilde{\mathbf{q}}, \mathbf{q}^\star)
  \Psi_{l_i{\kappa}}(\tilde{\mathbf{q}} \tilde{\nu})|^2
  \nonumber                                                             \\
                                                     & =
  \frac{1}{N^2} \sum_{i\kappa}
  \bigg| \sum_{j=1}^N
  \mathbf{W}_{T_j^{-1}(l_i){\kappa}}(\tilde{\mathbf{q}}\tilde{\nu})
  e^{i\mathbf{q}^\star \cdot \mathbf{R}_{l_j}} \bigg|^2
  \\
  \label{eq:phonon-unfolding-weight2}
                                                     & =
  \frac{1}{N} \sum_{i\kappa}
  \sum_{j=1}^N
  \mathbf{W}^*_{{l_i}{\kappa}}(\tilde{\mathbf{q}}{\nu})
  \mathbf{W}_{T_j^{-1}(l_i){\kappa}}(\tilde{\mathbf{q}}\tilde{\nu})
  e^{i\mathbf{q}^\star \cdot \mathbf{R}_{l_j}}.
\end{align}
It can easily be checked that $ \sum_{i=1}^N w(\tilde{\mathbf{q}}\tilde{\nu},
  \mathbf{q}^\star_i)  = 1$ from Eq.~\ref{eq:completeness-relation} and the unit
normalization of the phonon eigenvectors. Interestingly, this result does not
require phonon eigenvectors of the perfect crystal. The second equation is
implemented in the phonopy code though the third equation may be more
intuitive. $\mathbf{W}_{T_j^{-1}(l_i){\kappa}}(\tilde{\mathbf{q}}\tilde{\nu})$
is the phonon eigenvector elements belonging to the lattice point
$T_j^{-1}(l_i)$. The unfolding weight indicates its correlation with
$\mathbf{W}_{{l_i}{\kappa}}(\tilde{\mathbf{q}}{\nu})$ after the phase shift
$e^{i\mathbf{q}^\star \cdot \mathbf{R}_{l_j}}$. In the example of
Fig.~\ref{fig:unfolding-supercell}, the phase shifts are integer multiples of
$\pi$, since the 2D supercell model is made of $2 \times 2$ primitive cells.

\begin{figure}[ht]
  \begin{center}
    \includegraphics[width=0.9\linewidth]{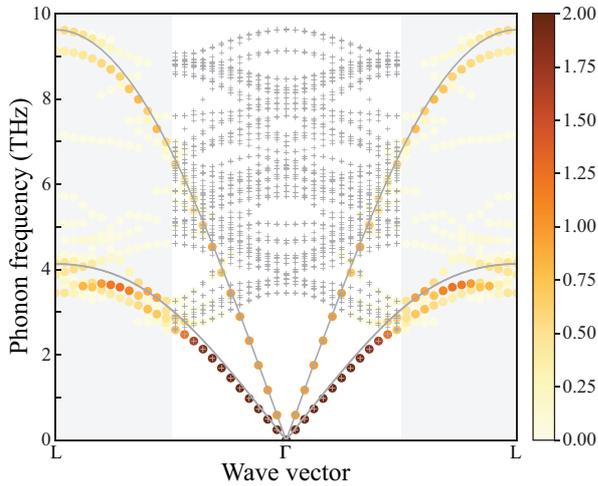}
    \caption{Phonon band unfolding of Al with a vacancy. Color scale indicates
      unfolding weights of phonon modes depicted by filled circle symbols, where
      those of the degenerated modes are summed. Phonon modes calculated for the
      defective supercell are presented by + (plus) symbols. Unshaded and shaded
      backgrounds show Brillouin zones of the supercell that belong to
      $\mathbf{q}^\star$ of $\Gamma$ and L, respectively. Only a part of
      weights are unfolded into these two commensurate points among 32
      commensurate points. Although this
      unfolding technique requires no reference states, for the comparison,
      phonon band structure of the perfect supercell is shown by solid curves.
      \label{fig:band-unfolding-VAl} }
  \end{center}
\end{figure}

An example of the phonon band unfolding technique applied to Al with a vacancy
is shown in Fig.~\ref{fig:band-unfolding-VAl}. The corresponding perfect
supercell (32 atoms) is constructed by $2 \times 2 \times 2$ of conventional
unit cell (4 atoms). The primitive cell contains 1 atom and
\begin{align}
  \boldsymbol{M}_{\text{p}\rightarrow\text{s}} = \begin{pmatrix*}[r]
                                                   -2 & 2 & 2 \\
                                                   2 & -2 & 2 \\
                                                   2 & 2 & -2
                                                 \end{pmatrix*}.
\end{align}

Under fixed basis vectors of the supercell with a vacancy, internal atomic
positions are relaxed. For the unfolding, $\mathbf{q}$ points are sampled along
the BZ path L--$\Gamma$--L of the primitive cell (see
Fig.~\ref{fig:band-unfolding-VAl}). Those points can be uniquely decomposed as
$\mathbf{q}=\tilde{\mathbf{q}} + \mathbf{q}^\star$, where $\tilde{\mathbf{q}}$
belongs to the BZ of the supercell, and  $\mathbf{q}^\star$ is a reciprocal
lattice vector of the supercell. For the above example, $\mathbf{q}^\star$ is
found to be $\Gamma$ or L, depending on the position of $\mathbf{q}$ along the
path L--$\Gamma$--L. Those two possibilities are represented using unshaded and
shaded backgrounds in Fig.~\ref{fig:band-unfolding-VAl}.

The phonon modes of the defective supercell are calculated at those
$\tilde{\mathbf{q}}$ points. The projection operator of
Eq.~(\ref{eq:unfolding-projector}) is $\mathbf{q}^\star$ dependent and only a
fraction of the phonon modes are unfolded along the L--$\Gamma$--L path, because
the selected $\mathbf{q}^\star$ points are only a subset of the commensurate
$\mathbf{q}^\star$ points in the summation of
Eq.~(\ref{eq:band-unfolding-decomposition}). In
Fig.~\ref{fig:band-unfolding-VAl}, though we can see phonon
frequencies are perturbed by the vacancy, the unfolded phonon band structure
roughly follows the phonon band structure of the perfect supercell.

Substitutional defects can be managed straightforwardly. Interstitial defects
can be handled in the same way as the vacancy case by defining corresponding
supercell differently.\cite{Moxon-unfolding-2022} Suppose we have one
intersticial defect in the supercell model. It is treated as a usual atom in the
primitive cell of the corresponding perfect supercell. In the defective
supercell, the interstitial atom is located in one primitive cell, and in the
other primitive cells, the interstitial sites are treated as vacancies.

\section{Conclusion}

A scientific simulation code implements mathematical models that are described
with formulae. In those formulae, the information is often not detailed enough
for direct computer implementation, and the detailed knowledge can be implicit.
Scientific software developers may have to learn it from skilled people around,
otherwise find it independently. To ease the latter situation in the modern era
of distributed software development, we aimed to provide details of the
computational methods employed in the phonopy and phono3py codes, keeping the
formalism as close as possible to the implementations.

Not all of the computational methods described in this review
are specific to the phonon calculation or the supercell approach. For example,
handling of the regular grids and the linear tetrahedron method presented are
applicable to electronic structure calculations. We therefore hope that this
review will be a useful source of information for software developers in
condensed matter science.

\section*{ACKNOWLEDGMENTS}
This work was supported by MEXT Japan through ESISM (Elements Strategy
Initiative for Structural Materials) of Kyoto University and JSPS KAKENHI Grant
Numbers JP21K04632 and JP21K03424. L. C. thanks the Institut Universitaire de
France and the EXPLOR center hosted by the University de Lorraine.

\appendix

\section{Functions used in the linear tetrahedron method \label{tetra_fc}} In
this section we give explicit formulae for  $g^i$, $n^i$, $I_k^i$, and
$J_k^i$.\cite{MacDonald-tetrahedron-1979} They assume different forms depending
on the position of $\omega$ in the sequence  $\omega^i_1 < \omega^i_2 <
  \omega^i_3 <\omega^i_4$. Also, they are expressed using $f_{nm}^i$ and
$\Delta_{nm}^i$ which are defined as
\begin{align}
   & f_{nm}^i \equiv (\omega - \omega_m^i) / (\omega_n^i - \omega_m^i),
  \;\; f_{nm}^i + f_{mn}^i = 1,                                         \\
   & \Delta_{nm}^i \equiv \omega_n^i - \omega_m^i.
\end{align}

$g^i$, $n^i$, $I^i$, and $J^i$ satisfy the following relations:
\begin{align}
  \label{eq:tetrahedron-n-to-g}
  \dv{n^i}{\omega}         & = g^i,       \\
  \label{eq:tetrahedron-J-to-I}
  \dv{(n^i J^i_k)}{\omega} & = g^i I^i_k.
\end{align}
It may be useful to remember
\begin{align}
  f_{n'm}^i /\Delta_{nm}^i & = f_{nm}^i /\Delta_{n'm}^i, \\
  \dv{f_{nm}^i}{\omega}    & = \frac{1}{\Delta_{nm}^i}.
\end{align}

\subsection{$\omega < \omega^i_1$}

\begin{align}
  g^i & = 0. \\ n^i &= 0. \\ I^i_k &= 0. \\ J^i_k &= 0.
\end{align}

\subsection{$\omega^i_1 < \omega < \omega^i_2$}

As shown in Fig.~\ref{fig:cut-tetrahedron} (a),  the volume to consider is a
tetrahedron with vertices located at $\mathbf{q}_1$, $\mathbf{q}_\alpha =
  f_{21}^i\mathbf{q}_2 + f_{12}^i\mathbf{q}_1$, $\mathbf{q}_\beta =
  f_{31}^i\mathbf{q}_3 + f_{13}^i\mathbf{q}_1$, $\mathbf{q}_\gamma =
  f_{41}^i\mathbf{q}_4 + f_{14}^i\mathbf{q}_1$, and
\begin{align}
  n^i = & f_{21}^i f_{31}^i f_{41}^i.
\end{align}
Inside this volume, $F(\mathbf{q})$ is linearly interpolated, i.e.,
$F(\mathbf{q}_\alpha) \approx f_{21}^i F(\mathbf{q}_2) + f_{12}^i
  F(\mathbf{q}_1)$, $\ldots$, and the contribution of $i$-th
tetrahedron to $J(\omega)$ is approximated as $V_\mathrm{t} n^i [F(\mathbf{q}_1)
  + F(\mathbf{q}_\alpha) + F(\mathbf{q}_\beta) + F(\mathbf{q}_\gamma)]/4$. By
substituting it in Eq.~(\ref{eq:tetrahedron-J-omega}), we obtain
\begin{align}
  J^i_1 = &
  (1 + f_{12}^i + f_{13}^i + f_{14}^i)/4.     \\
  J^i_k = & f_{k1}^i / 4, \;\;\; k = 2, 3, 4.
\end{align}
Using Eqs.~(\ref{eq:tetrahedron-n-to-g}) and (\ref{eq:tetrahedron-J-to-I}),
\begin{align}
  g^i   & = 3n^i / (\omega - \omega_1^i) = 3 f_{21}^i f_{31}^i f_{41}^i
  / (\omega - \omega_1^i)
  \nonumber                                                             \\
        & = 3 f_{21}^i f_{31}^i / \Delta_{41}^i.                        \\
  I^i_1 & = (f_{12}^i + f_{13}^i + f_{14}^i)/3.                         \\
  I^i_k & = f_{k1}^i / 3, \;\;\; k = 2, 3, 4.
\end{align}

\subsection{$\omega^i_2 < \omega < \omega^i_3$}
In this case, the occupied part of the tetrahedron is the sum of the following
three tetrahedra. Vertices of the first tetrahedron are $\mathbf{q}_1$,
$\mathbf{q}_2$, $\mathbf{q}_\alpha = f_{42}^i \mathbf{q}_4 + f_{24}^i
  \mathbf{q}_2$, $\mathbf{q}_\beta = f_{32}^i \mathbf{q}_3 + f_{23}^i
  \mathbf{q}_2$, and its volume is $V_{12\alpha\beta}^i = V_\mathrm{t} f_{42}^i
  f_{32}^i$.
The vertices of the second tetrahedron are $\mathbf{q}_1$, $\mathbf{q}_\alpha$,
$\mathbf{q}_\beta$, $\mathbf{q}_\delta = f_{41}^i \mathbf{q}_4 + f_{14}^i
  \mathbf{q}_1$, and its volume  is $V_{1\alpha\beta\delta}^i =
  V_\mathrm{t}f_{41}^i f_{24}^i f_{32}^i$.
The vertices of the third tetrahedron are $\mathbf{q}_1$, $\mathbf{q}_\beta$,
$\mathbf{q}_\delta$, $\mathbf{q}_\gamma = f_{31}^i \mathbf{q}_3 + f_{13}^i
  \mathbf{q}_1$, and its volume is $V_{1\beta\delta\gamma}^i = V_\mathrm{t}
  f_{41}^i f_{31}^i f_{23}^i$. The locations of the vertices are depicted in
Fig.~\ref{fig:cut-tetrahedron} (b). We obtain
\begin{align}
  n^i =   & (V_{12\alpha\beta}^i + V_{1\alpha\beta\delta}^i +
  V_{1\beta\delta\gamma}^i) / V_\mathrm{t}
  \nonumber
  \\
  =       & f_{42}^i f_{32}^i + f_{41}^i f_{24}^i f_{32}^i + f_{41}^i f_{31}^i
  f_{23}^i.
  \\
  J^i_1 = & [V_{12\alpha\beta}^i + V_{1\alpha\beta\delta}^i (1 + f_{14}^i)
      \nonumber
  \\
          & + V_{1\beta\delta\gamma}^i (1 + f_{14}^i + f_{13}^i)] /
  (4 n^i V_\mathrm{t}).
  \\
  J^i_2 = & [V_{12\alpha\beta}^i (1 + f_{24}^i + f_{23}^i) +
      V_{1\alpha\beta\delta}^i(f_{24}^i + f_{23}^i)
      \nonumber
  \\
          & + V_{1\beta\delta\gamma}^i f_{23}^i] / (4 n^i V_\mathrm{t}).
  \\
  J^i_3 = & [V_{12\alpha\beta}^i f_{32}^i + V_{1\alpha\beta\delta}^i
      f_{32}^i + V_{1\beta\delta\gamma}^i (f_{32}^i + f_{31}^i)]
  / (4 n^i V_\mathrm{t}).
  \\
  J^i_4 = & [V_{12\alpha\beta}^i f_{42}^i + V_{1\alpha\beta\delta}^i
      (f_{42}^i + f_{41}^i) + V_{1\beta\delta\gamma}^i f_{41}^i]
  / (4 n^i V_\mathrm{t}).
  \\
  g^i =   & 3(f_{23}^i f_{31}^i + f_{32}^i f_{24}^i) / \Delta^i_{41} .
  \\
  I^i_1 = & f_{14}^i/3 + f_{13}^i f_{31}^i f_{23}^i / (g^i \Delta_{41}^i).
  \\
  I^i_2 = & f_{23}^i/3 + (f_{24}^i)^2 f_{32}^i / (g^i \Delta_{41}^i).
  \\
  I^i_3 = & f_{32}^i/3 + (f_{31}^i)^2 f_{23}^i / (g^i \Delta_{41}^i).
  \\
  I^i_4 = & f_{41}^i/3 + f_{42}^i f_{24}^i f_{32}^i / (g^i \Delta_{41}^i).
\end{align}

\subsection{$\omega^i_3 < \omega < \omega^i_4$}
As shown in Fig.~\ref{fig:cut-tetrahedron} (c), the occupied part of the
tetrahedron is the full tetrahedron minus the tetrahedron with vertices at
$\mathbf{q}_4$, $\mathbf{q}_\beta = f_{42}^i\mathbf{q}_4 +
  f_{24}^i\mathbf{q}_2$, $\mathbf{q}_\delta = f_{41}^i\mathbf{q}_4 +
  f_{14}^i\mathbf{q}_1$ and $\mathbf{q}_\gamma = f_{43}^i\mathbf{q}_4 +
  f_{34}^i\mathbf{q}_3$. Its  volume is $V_{4\beta\delta\gamma}^i =
  V_\mathrm{t}f_{14}^i f_{24}^i f_{34}^i$. Therefore, the
contribution of $i$-th tetrahedron to $J(\omega)$ is approximated as
$\{V_\mathrm{t}[F(\mathbf{q}_1) + F(\mathbf{q}_2) + F(\mathbf{q}_3) +
    F(\mathbf{q}_4)] - V_{4\beta\delta\gamma}^i [F(\mathbf{q}_4) +
  F(\mathbf{q}_\beta) + F(\mathbf{q}_\gamma) + F(\mathbf{q}_\delta)] \}/4$.
We obtain
\begin{align}
  n^i =   & (1 - f_{14}^i f_{24}^i f_{34}^i).              \\
  J^i_1 = & [1 - (f_{14}^i)^2 f_{24}^i f_{34}^i] / 4n^i.   \\
  J^i_2 = & [1 - f_{14}^i (f_{24}^i)^2 f_{34}^i] / 4n^i.   \\
  J^i_3 = & [1 - f_{14}^i f_{24}^i (f_{34}^i)^2] / 4n^i.   \\
  J^i_4 = & [1 - f_{14}^i f_{24}^i f_{34}^i(1 + f_{41}^i +
  f_{42}^i + f_{43}^i)] / 4n^i.                            \\
  g^i =   & 3(1 - n^i)/(\omega_4^i - \omega) = 3 f_{24}^i
  f_{34}^i / \Delta_{41}^i.                                \\
  I^i_k = & f_{k4}^i / 3, \;\;\; k = 1, 2, 3.              \\
  I^i_4 = & (f_{41}^i + f_{42}^i + f_{43}^i) / 3.
\end{align}

\subsection{$\omega^i_4 < \omega$}
The tetrahedron is fully occupied, therefore,
\begin{align}
  g^i & = 0. \\ n^i & = 1. \\ I^i_k &= 0. \\ J^i_k &= 1/4.
\end{align}

\section{Summation formulae \label{sum_f}}
When the supercell lattice vectors are collinear to those of the primitive cell,
\begin{align}
  (\mathbf{a}_\text{s},
  \mathbf{b}_\text{s}, \mathbf{c}_\text{s}) =
  (\mathbf{a}_\text{p},
  \mathbf{b}_\text{p}, \mathbf{c}_\text{p})
  \boldsymbol{M}_{\text{p}\rightarrow\text{s}}.
\end{align}
With $ \boldsymbol{M}_{\text{p}\rightarrow\text{s}}= \text{diag}(n_1,n_2,n_3)$,
it is proved in most textbook that the set of lattice vectors
\begin{align}
  \mathbf{R}_l & = (\mathbf{a}_\text{p},  \mathbf{b}_\text{p}, \mathbf{c}_\text{p})
  \begin{pmatrix} l_1\\l_2 \\l_3\end{pmatrix}, \quad l_i=0,\ldots,n_i-1
  \\
               & = (\mathbf{a}_\text{p},  \mathbf{b}_\text{p}, \mathbf{c}_\text{p})
  \boldsymbol{l}
\end{align}
and their conjugated wave vectors
\begin{align}
  \mathbf{q} & =
  (\mathbf{a}_\text{p}^{*}, \mathbf{b}_\text{p}^{*}, \mathbf{c}_\text{p}^{*})
  \begin{pmatrix} p_1/n_1\\p_2/n_2 \\p_3/n_3\end{pmatrix},
  \quad p_i=0,\ldots,n_i-1                   \\
             & =
  (\mathbf{a}_\text{s}^{*},  \mathbf{b}_\text{s}^{*}, \mathbf{c}_\text{s}^{*})
  \begin{pmatrix} p_1\\p_2\\p_3\end{pmatrix} \\
             & =
  (\mathbf{a}_\text{s}^{*},  \mathbf{b}_\text{s}^{*}, \mathbf{c}_\text{s}^{*})
  \boldsymbol{p},
\end{align}
fulfill the summation formulae
\begin{align}
   & \frac{1}{\det(\boldsymbol{M}_{\text{p}\rightarrow\text{s}})}
  \sum_{\{\mathbf{R}_l \}} e^{i \mathbf{q}\cdot \mathbf{R}_l} =
  \delta_{\mathbf{q},\mathbf{G}}, \label{eq:summation-formula-1}  \\
   & \frac{1}{\det(\boldsymbol{M}_{\text{p}\rightarrow\text{s}})}
  \sum_{\{\mathbf{q} \}} e^{-i \mathbf{q}\cdot \mathbf{R}_l} =
  \delta_{\mathbf{R}_l,\mathbf{R}_L}, \label{eq:summation-formula-2}
\end{align}
where $\mathbf{G}$ is any reciprocal lattice vector, and $\mathbf{R}_L$ any
supercell lattice vector.

When the supercell is not collinear to the primitive cell, the set of lattice
vectors, $\{\mathbf{R}_l\}$, are those integer linear combinations of
$(\mathbf{a}_\text{p}, \mathbf{b}_\text{p}, \mathbf{c}_\text{p})$ located within
the supercell $ (\mathbf{a}_\text{s}, \mathbf{b}_\text{s},
  \mathbf{c}_\text{s})$, and the set of wave vectors, $\{\mathbf{q}\}$, are those
integer linear combinations of $ (\mathbf{a}_\text{s}^{*},
  \mathbf{b}_\text{s}^{*}, \mathbf{c}_\text{s}^{*})$ located within the reciprocal
cell $(\mathbf{a}_\text{p}^{*},  \mathbf{b}_\text{p}^{*},
  \mathbf{c}_\text{p}^{*})$. Since the above summations can not easily be split
into the product of three geometric series, it is not obvious whether the above
results are still valid.

However, we have seen in Sec.~\ref{seq:supercell-construction} that SNF like
transformations can always be used to obtain primitive cell and supercell which
are collinear, and for which the summation formulae
(\ref{eq:summation-formula-1}) and (\ref{eq:summation-formula-2}) are obviously
valid. We have
\begin{align}
   & (\tilde{\mathbf{a}}_\text{p}, \tilde{\mathbf{b}}_\text{p},
  \tilde{\mathbf{c}}_\text{p}) =
  (\mathbf{a}_\text{p},  \mathbf{b}_\text{p}, \mathbf{c}_\text{p})
  \boldsymbol{P}^{-1},
  \\
   & (\tilde{\mathbf{a}}_\text{s}, \tilde{\mathbf{b}}_\text{s},
  \tilde{\mathbf{c}}_\text{s}) =
  (\mathbf{a}_\text{s},  \mathbf{b}_\text{s}, \mathbf{c}_\text{s})
  \boldsymbol{Q},
\end{align}
and therefore
\begin{align}
  (\tilde{\mathbf{a}}_\text{p}^{*}, \tilde{\mathbf{b}}_\text{p}^{*},
  \tilde{\mathbf{c}}_\text{p}^{*}) & =
  (\mathbf{a}_\text{p}^{*}, \mathbf{b}_\text{p}^{*}, \mathbf{c}_\text{p}^{*})
  \boldsymbol{P}^{\intercal}, \label{eq:unimodular-transformation-1}
  \\
  (\tilde{\mathbf{a}}_\text{s}^{*}, \tilde{\mathbf{b}}_\text{s}^{*},
  \tilde{\mathbf{c}}_\text{s}^{*}) & =
  (\mathbf{a}_\text{s}^{*},  \mathbf{b}_\text{s}^{*}, \mathbf{c}_\text{s}^{*})
  \boldsymbol{Q}^{-\intercal} \label{eq:unimodular-transformation-2}
  \\
                                   & =
  (\tilde{\mathbf{a}}_\text{p}^{*}, \tilde{\mathbf{b}}_\text{p}^{*},
  \tilde{\mathbf{c}}_\text{p}^{*}) \boldsymbol{D}^{-1},
\end{align}
while the set of lattice vectors and wave vectors,
\begin{align}
   & \tilde{\mathbf{R}}_l = (\tilde{\mathbf{a}}_\text{p},
  \tilde{\mathbf{b}}_\text{p}, \tilde{\mathbf{c}}_\text{p})
  \tilde{\boldsymbol{l}},                                     \\
   & \tilde{ \mathbf{q} } = (\tilde{\mathbf{a}}_\text{s}^{*},
  \tilde{\mathbf{b}}_\text{s}^{*}, \tilde{\mathbf{c}}_\text{s}^{*})
  \tilde{\boldsymbol{p}},
\end{align}
with $ \tilde{l}_i=0,\ldots,D_i-1$ and $\tilde{p}_i=0,\ldots,D_i-1$, fulfills,
\begin{align}
   & \frac{1}{\det(\boldsymbol{D})} \sum_{\{ \tilde{\mathbf{R}}_l \}}
  e^{i \tilde{\mathbf{q}}\cdot \tilde{\mathbf{R}}_l} =
  \delta_{\tilde{\mathbf{q}},\mathbf{G}},                             \\
   & \frac{1}{\det(\boldsymbol{D})} \sum_{\{ \tilde{ \mathbf{q}} \}}
  e^{-i \tilde{\mathbf{q}}\cdot \tilde{\mathbf{R}}_l} =
  \delta_{\tilde{\mathbf{R}}_l,\mathbf{R}_L}.
\end{align}

Since the same lattices are generated by the unimodular transformations
(\ref{eq:unimodular-transformation-1}) and
(\ref{eq:unimodular-transformation-2}), we can write
\begin{align}
   & \tilde{\mathbf{R}}_l = \mathbf{R}_l + \mathbf{R}_L, \\
   & \tilde{\mathbf{q}} = \mathbf{q} + \mathbf{G}
\end{align}
where $ \mathbf{R}_L$ is a supercell lattice vector used to bring $
  \tilde{\mathbf{R}}_l$ within the original supercell $(\mathbf{a}_\text{s},
  \mathbf{b}_\text{s}, \mathbf{c}_\text{s})$,  and $\mathbf{G}$ a reciprocal
lattice vector used to bring $ \tilde{ \mathbf{q} } $ to the original
reciprocal cell  $ (\mathbf{a}_\text{p}^{*}, \mathbf{b}_\text{p}^{*},
  \mathbf{c}_\text{p}^{*})$. Since $\mathbf{q}\cdot \mathbf{R}_L$,
$\mathbf{G}\cdot \mathbf{R}_L$, $\mathbf{G}\cdot \mathbf{R}_l$ are multiple of
$2 \pi$, and
$\det(\boldsymbol{M}_{\text{p}\rightarrow\text{s}})=\det(\boldsymbol{D})$, we
obtain Eqs.~(\ref{eq:summation-formula-1}) and (\ref{eq:summation-formula-2})
for arbitrary supercell and primitive cell related by an integer matrix
$\boldsymbol{M}_{\text{p}\rightarrow\text{s}}$. Notice however that the
definition of the set of lattice vectors and wave vectors for which those
summation formulae hold is now more general. The wave vectors are along the
reciprocal vectors of the supercell and located within the reciprocal cell of
the primitive lattice, while the lattice vectors are along the primitive
lattice vectors and located within the supercell.

\section{Projection into primitive cell Bloch states \label{PC_Bloch}}
From the definition of the projection operator,
\begin{widetext}
  \begin{align}
    \hat{T}_i \hat{P}(\tilde{\mathbf{q}}, \mathbf{q^\star})
    \Psi_{l{\kappa}}(\tilde{\mathbf{q}} \tilde{\nu}) & =
    \hat{T}_i  \frac{1}{N} \sum_{j=1}^N
    e^{i (\tilde{\mathbf{q}} + \mathbf{q^\star}) \cdot \mathbf{R}_{l_j}} \hat{T}_j
    \Psi_{l{\kappa}}(\tilde{\mathbf{q}} \tilde{\nu})                                                         \\
                                                     & = e^{-i (\tilde{\mathbf{q}} + \mathbf{q^\star})
        \cdot \mathbf{R}_{l_i}}  \frac{1}{N} \sum_{j=1}^N
    e^{i (\tilde{\mathbf{q}} + \mathbf{q^\star})
        \cdot (\mathbf{R}_{l_j}+\mathbf{R}_{l_i})} \hat{T}_i \hat{T}_j
    \Psi_{l{\kappa}}(\tilde{\mathbf{q}} \tilde{\nu})                                                         \\
                                                     & = e^{-i (\tilde{\mathbf{q}} + \mathbf{q^\star})
        \cdot \mathbf{R}_{l_i}}  \frac{1}{N} \sum_{j=1}^N
    e^{i (\tilde{\mathbf{q}} + \mathbf{q^\star}) \cdot
        (\mathbf{R}_{l_j}+\mathbf{R}_{l_i})} e^{i\tilde{\mathbf{q}} \cdot
        (\mathbf{R}_{l{\kappa}} - \mathbf{R}_{l_j}-\mathbf{R}_{l_i})}
    \mathbf{W}_{(T_iT_j)^{-1}(l)\kappa}(\tilde{\mathbf{q}} \tilde{\nu})                                      \\
                                                     & = e^{-i (\tilde{\mathbf{q}} + \mathbf{q^\star}) \cdot
        \mathbf{R}_{l_i}} \frac{1}{N} \sum_{j=1}^N
    e^{i  \mathbf{q^\star} \cdot (\mathbf{R}_{l_j}+\mathbf{R}_{l_i})}
    e^{i\tilde{\mathbf{q}} \cdot \mathbf{R}_{l{\kappa}} }
    \mathbf{W}_{(T_iT_j)^{-1}(l)\kappa}(\tilde{\mathbf{q}} \tilde{\nu}).
  \end{align}
  $\mathbf{R}_{l_j}+\mathbf{R}_{l_i}$ may be outside the supercell, therefore we
  write $\mathbf{R}_{l_j}+\mathbf{R}_{l_i}=\mathbf{R}_{l_k}+\mathbf{R}_L$, where
  $\mathbf{R}_L$ is a supercell lattice vector. Because $\mathbf{W}_{l
      \kappa}(\tilde{\mathbf{q}} \tilde{\nu})$ has the periodicity of the supercell,
  and $e^{i  \mathbf{q^\star} \cdot \mathbf{R}_L}=1$, we obtain
  \begin{align}
    \hat{T}_i \hat{P}(\tilde{\mathbf{q}}, \mathbf{q^\star})
    \Psi_{l{\kappa}}(\tilde{\mathbf{q}} \tilde{\nu})
     & = e^{-i (\tilde{\mathbf{q}} + \mathbf{q^\star})
        \cdot \mathbf{R}_{l_i}} \frac{1}{N} \sum_{k=1}^N
    e^{i  \mathbf{q^\star} \cdot \mathbf{R}_{l_k}} e^{i\tilde{\mathbf{q}}
    \cdot \mathbf{R}_{l{\kappa}} }
    \mathbf{W}_{T_k^{-1}(l)\kappa}(\tilde{\mathbf{q}} \tilde{\nu})             \\
     & = e^{-i (\tilde{\mathbf{q}} + \mathbf{q^\star}) \cdot\mathbf{R}_{l_i}}
    \frac{1}{N} \sum_{k=1}^N e^{i (\tilde{\mathbf{q}}+\mathbf{q^\star})
        \cdot \mathbf{R}_{l_k}}
    e^{i\tilde{\mathbf{q}} \cdot  (\mathbf{R}_{l{\kappa}}-\mathbf{R}_{l_k}) }
    \mathbf{W}_{T_k^{-1}(l)\kappa}(\tilde{\mathbf{q}} \tilde{\nu})             \\
     & =  e^{-i (\tilde{\mathbf{q}} + \mathbf{q^\star}) \cdot\mathbf{R}_{l_i}}
    \frac{1}{N} \sum_{k=1}^N
    e^{i (\tilde{\mathbf{q}}+\mathbf{q^\star}) \cdot \mathbf{R}_{l_k}} \hat{T}_k
    \Psi_{l{\kappa}}(\tilde{\mathbf{q}} \tilde{\nu})                           \\
     & =e^{-i (\tilde{\mathbf{q}} + \mathbf{q^\star}) \cdot
    \mathbf{R}_{l_i}} \hat{P}(\tilde{\mathbf{q}}, \mathbf{q^\star})
    \Psi_{l{\kappa}}(\tilde{\mathbf{q}} \tilde{\nu}).
  \end{align}

\end{widetext}

\bibliography{JPCM}
\end{document}